\documentclass[letterpaper,12pt]{article}

\pdfoutput=1
\usepackage{graphicx}
\usepackage{amsmath}
\usepackage{amsfonts}
\usepackage{hyperref}
\usepackage{bm}

\makeatletter
\renewcommand\section{\@startsection {section}{1}{\z@}%
{-3.5ex \@plus -1ex \@minus -0.2ex}%
{2.3ex \@plus 0.2ex}%
{\normalfont\normalsize\bfseries}}

\renewcommand\subsection{\@startsection{subsection}{2}{\z@}%
{-3.25ex \@plus -1ex \@minus -0.2ex}%
{1.5ex \@plus 0.2ex}%
{\normalfont\normalsize\bfseries}}

\def\@seccntformat#1{\csname the#1\endcsname.\quad}
\makeatother

\newcommand{\oline}[1]{\overline{\mkern-1.0mu#1\mkern0.0mu}}
\newcommand\redots{\makebox[0.85em][c]{.\hfil.\hfil.}}
\newcommand{\overbar}[1]{\mkern 4.3mu\overline{\mkern-4.3mu#1\mkern-1.5mu}\mkern 1.5mu}
\newcommand\pvalue{\mbox{\footnotesize{$\mathcal{J}$}}}
\newcommand\Fprime{F\hspace{0.05em}^{\prime}}

\begin{document}

\setlength{\baselineskip}{4.5ex}

\noindent
{\bf \LARGE Integrated organic inference (IOI):}\\[2ex]
{\bf \LARGE A reconciliation of statistical paradigms}\\[3ex]

\noindent
{\bf Russell J. Bowater}\\
\emph{Independent researcher, Sartre 47, Acatlima, Huajuapan de Le\'{o}n, Oaxaca, C.P.\ 69004,
Mexico. Email address: as given on arXiv.org. Twitter profile:
\href{https://twitter.com/naked_statist}{@naked\_statist}\\ Personal website:
\href{https://sites.google.com/site/bowaterfospage}{sites.google.com/site/bowaterfospage}}
\\[2ex]

\noindent
{\small \bf Abstract:}
{\small
It is recognised that the Bayesian approach to inference can not adequately cope with all the types
of pre-data beliefs about population quantities of interest that are commonly held in practice. In
particular, it generally encounters difficulty when there is a lack of such beliefs over some or
all the parameters of a model, or within certain partitions of the parameter space concerned. To
address this issue, a fairly comprehensive theory of inference is put forward called integrated
organic inference that is based on a fusion of Fisherian and Bayesian reasoning. Depending on the
pre-data knowledge that is held about any given model parameter, inferences are made about the
parameter conditional on all other parameters using one of three methods of inference, namely
organic fiducial inference, bispatial inference and Bayesian inference. The full conditional
post-data densities that result from doing this are then combined using a framework that allows a
joint post-data density for all the parameters to be sensibly formed without requiring these full
conditional densities to be compatible. Various examples of the application of this theory are
presented. Finally, the theory is defended against possible criticisms partially in terms of what
was previously defined as generalised subjective probability.}
\\[3ex]

\pagebreak
\section{Introduction}

The general problem of making inferences about a population on the basis of a small random sample
from that population has long been of great interest to scientific researchers.
This problem is often addressed by making the assumption that, in the population, the distribution
of the measurements being considered is a member of a given parametric family of distributions.
Although this assumption can be criticised, we will choose in this paper to examine problems of
inference that are constrained by this assumption.
Our justification for this is that, first, this class of problems has substantial importance in
its own right, and second, resolving such problems can be viewed as a convenient first step towards
tackling cases in which making such an assumption is not appropriate.
Therefore, let us suppose that the data set to be analysed $x=\{x_1, x_2, \ldots, x_n\}$ was drawn
from a joint density or mass function $g(x\,|\,\theta)$ that depends on a set of parameters
$\theta =\{\theta_i : i=1,2,\ldots,k\}$, where each $\theta_i$ is a one-dimensional variable.

A way of classifying the nature of the problem that is encountered in trying to make inferences
about the set of parameters $\theta$ is to do so on the basis of the type of knowledge that was
held about these parameters before the data were observed.
In this respect, it can be argued that the three most common types of pre-data opinion that, in
practice, are naturally held about any given model parameter $\theta_j$ conditional on all other
parameters $\theta_{-j}=\{ \theta_1,\ldots,\theta_{j-1},\theta_{j+1},\dots, \theta_{k} \}$ being
known are as follows:

\vspace{1.5ex}
\noindent
1) Nothing or very little is known about the parameter.

\vspace{1.5ex}
\noindent
2) It is felt that the parameter may well be close to a specific value, which may for example
indicate the absence of a treatment effect, or the lack of a correlation between variables, but
apart from this nothing or very little is known about the parameter.
Some examples of where it would be reasonable to hold this type of pre-data opinion were given in
Bowater~(2019b).

\pagebreak
\noindent
3) We know enough about the parameter for our opinion about it to be satisfactorily represented by
a probability density or mass function over the parameter.

\vspace{1.5ex}
For the reason just given, each of these types of pre-data opinion about the parameter $\theta_j$
will therefore be treated as corresponding to a distinct problem of inference.
Nevertheless, since our pre-data opinions about each of the parameters in any given set of
parameters $\theta$ may well fall into different categories among the three being considered, it
may be necessary to address two or all three of these types of problem in any particular scenario.

These problems are the three problems of inference that will be of principal interest in what
follows. More specifically, the aim of the present paper will be to show how these problems can be
dealt with in a harmonious manner by using an approach to inference based on a fusion of Fisherian
(as attributed to R.\ A.\ Fisher) and Bayesian reasoning.
Of course, given the obvious incompatibilities that exist between, and to some extent even within,
these two schools of reasoning, we will need to be given some liberty in how each of these
approaches to inference is interpreted.

In this respect, although the theory that will be outlined is based on a type of probability that
is inherently subjective, and therefore not frequentist as in the Fisherian paradigm, it is not the
same type of probability that is commonly regarded as underlying subjective Bayesian theory.
Instead, it is a generalised form of subjective probability that effectively allows probability
distributions to be distinguished according to where they are on a scale that goes from them being
virtually objective to them being extremely subjective.
This type of probability was referred to as generalised subjective probability in Bowater~(2018b).
Furthermore, the theory to be presented relies on various concepts that are heavily used by
frequentist statisticians, e.g.\ sufficient and ancillary statistics, point estimators and their
distributions, the classical \pagebreak notion of significance, and also one very important idea
that during his own lifetime was chiefly advocated by Fisher himself, namely the fiducial argument.
We are not suggesting, though, that the proposed methodology should be judged positively simply
because it represents a compromise between competing schools of inference, rather we recommend,
quite naturally, that it should be evaluated on the basis of its effectiveness in dealing with the
particular inferential challenge that has been set out.

To give a little more detail, each of the three aforementioned problems of inference will be
addressed using a method that is specific to the problem concerned, and although this results in
the use of three methods that are of a clearly different nature, these methods are nevertheless
compatible with the overall framework of inference that will be put forward. In particular, the
first type of problem will be tackled using what, in Bowater~(2019a), was called organic fiducial
inference. On the other hand, the second problem will be addressed using what, in Bowater~(2019b),
was called bispatial inference. Finally, the third problem will be dealt with using Bayesian
inference. The overall framework just referred to provides a way of coordinating these distinct
methods of inference so that it is possible to simultaneously make inferences about all of the
parameters in the model.

Let us now briefly describe the structure of the paper. In the next five sections, we will present
summaries of the fundamental concepts and methods that form the basis of the general theory in
question, which will be called integrated organic inference (IOI). In particular, in the next two
sections we will summarise the theory of generalised subjective probability and the overall
framework of integrated organic inference.
Furthermore, after clarifying in Section~\ref{sec3} the interpretation that will be adopted in this
paper of the Bayesian approach to inference, concise accounts of the methods of organic fiducial
inference and bispatial inference will be given in Sections~\ref{sec4} and~\ref{sec5}.
Various examples of the application of integrated organic inference will then be outlined in detail
in Sections~\ref{sec6} to~\ref{sec7}.
In the final section of the paper (Section~\ref{sec8}), a discussion of this theory of inference
will be presented in the form of answers to questions that would be expected to naturally arise
about the theory when it is first evaluated.

The theory will be referred to as integrated organic inference (IOI) because it integrates what are
often considered to be conflicting approaches to inference into an overall framework that relies,
in general, on what can be viewed as being an organic simulation algorithm.
Furthermore, the type of inferences that this theory facilitates may, depending on the
circumstances, be regarded as being objective or very subjective, but are nevertheless always
organic, in the sense that they are intended to be only really understood by living subjects, e.g.\
humans, rather than primitive robots.

\vspace{3ex}
\section{Fundamental concepts and methods}
\label{sec9}

\vspace{1ex}
\subsection{Generalised subjective probability}
\label{sec1}

\vspace{1ex}
\noindent
{\bf Overview}

\vspace{1ex}
\noindent
Under this definition of probability, a probability distribution is defined by its (cumulative)
distribution function and the strength of this function relative to other distribution functions of
interest. The distribution function is defined as having the standard mathematical properties of
such a function.
Let us now briefly outline the notion of the strength of a distribution function and some of the
concepts that underlie this notion. Further details and examples of these concepts and of the
notion of strength itself can be found in Bowater~(2018b).

\vspace{3ex}
\noindent
{\bf Similarity}

\vspace{1ex}
\noindent
As in the aforementioned paper, let $S(A,B)$ denote the similarity that a given individual feels
there is between his confidence (or conviction) that an event $A$ will occur and his confidence (or
conviction) that an event $B$ will occur. For any three events $A$, $B$ and $C$, it is assumed that
an individual is capable of deciding whether or not the orderings $S(A,B) > S(A,C)$ and
$S(A,B) < S(A,C)$ are applicable. The notation $S(A,B) = S(A,C)$ is used to represent the case
where neither of these orderings apply.

\vspace{3ex}
\noindent
{\bf Reference set of events}

\vspace{1ex}
\noindent
Let $O=\{ O_1, O_2, \ldots, O_m \}$ be a finite ordered set of $m$ events that are mutually
exclusive and exhaustive. Also, let us assume that if $O(1)$, $O(2)$ and $O(3)$ are three subsets
of the set $O$ that contain the same number of events, then the following is true:
\vspace{1.5ex}
\[
S \left( \bigcup_{O_{\hspace{-0.05em}j} \in\hspace{0.05em} O(1)} O_{\hspace{-0.05em}j},
\bigcup_{O_{\hspace{-0.05em}j} \in\hspace{0.05em} O(2)} O_{\hspace{-0.05em}j} \right)
=\, S \left( \bigcup_{O_{\hspace{-0.05em}j} \in\hspace{0.05em} O(1)} O_{\hspace{-0.05em}j},
\bigcup_{O_{\hspace{-0.05em}j} \in\hspace{0.05em} O(3)} O_{\hspace{-0.05em}j} \right)
\vspace{2.5ex}
\]
for all possible choices of the subsets $O(1)$, $O(2)$ and $O(3)$.
Under this assumption, a reference set of events $R$ can be defined as follows:
\begin{equation}
\label{equ1}
R = \{ R(\lambda): \lambda \in \Lambda \}
\end{equation}
where $R(\lambda) = O_1 \cup O_2 \cup \cdots \cup O_{\lambda m}$ and
$\Lambda = \{ 1/m,\, 2/m,\, \ldots, (m-1)/m \}$.
For example, it should be clear that any given individual could easily decide that the set of all
the outcomes of randomly drawing a ball out of an urn containing $m$ distinctly labelled balls
could be the set $O$.

Equation~(\ref{equ1}) gives the definition of a reference set of events assuming that this set is
discrete. For the definition of a continuous reference set of events, see Bowater~(2018b).

\vspace{3ex}
\noindent
{\bf External strength of a distribution function}

\vspace{1ex}
\noindent
Let two continuous random variables $X$ and $Y$ of possibly different dimensions have elicited or
given distribution functions $F_X(x)$ and $G_Y(y)$ \pagebreak respectively. Also, we will specify
the set of events $\mathcal{F}[a]$ as follows:
\vspace{2ex}
\[
\mathcal{F}[a] = \left\{ \{ X \in \mathcal{A} \}: \int_{\mathcal{A}} f_X(x) dx = a \right\}\ \ \
\mbox{for $a \in [0,1]$}
\vspace{2.5ex}
\]
where $\{ X \in \mathcal{A} \}$ is the event that $X$ lies in the set $\mathcal{A}$ and $f_X(x)$ is
the density function corresponding to $F_X(x)$, and we will specify the set $\mathcal{G}[a]$ in the
same way but with respect to the variable $Y$ instead of the variable $X$ and the distribution
function $G_Y(y)$ instead of $F_X(x)$.

For a given discrete or continuous reference set of events $R$, we will now define the function
$F_X(x)$ as being externally stronger than the function $G_Y(y)$ at the resolution $\lambda$, where
$\lambda \in \Lambda$, if
\vspace{1.5ex}
\[
\underset{\mbox{\footnotesize $A\hspace{-0.1em} \in\hspace{-0.1em}
\mathcal{F}[\lambda]$}}{\min}\, S( A, R(\lambda) ) >
\underset{\mbox{\footnotesize $A\hspace{-0.1em} \in\hspace{-0.1em}
\mathcal{G}[\lambda]$}}{\max}\, S( A, R(\lambda) )
\vspace{3ex}
\]
An interpretation that could be given to this definition is that, if a particular individual judges
a function $F_X(x)$ as being externally stronger than a function $G_Y(y)$ then, relative to the
reference event $R(\lambda)$, the function $F_X(x)$ could be regarded as representing his
uncertainty about the variable $X$ better than $G_Y(y)$ represents his uncertainty about the
variable $Y$.

A definition of the internal rather than the external strength of a distribution function, and
other definitions of the external strength of a distribution function that are applicable to
discrete distribution functions and to distribution functions derived by formal systems of
reasoning, e.g.\ derived by applying the standard rules of probability, can be found in
Bowater~(2018b).

\pagebreak
\vspace{3ex}
\subsection{Overall framework of the theory}
\label{sec2}

\vspace{1ex}
\noindent
{\bf Brief outline}

\vspace{1ex}
\noindent
The general aim of the theory to be presented is to construct a joint density/mass function of all
the model parameters $\theta$ that accurately represents what is known about these parameters after
the data have been observed, i.e.\ what can be referred to as a post-data density function of these
parameters. Let this density function be denoted as $p(\theta\,|\,x)$. \linebreak To be more
specific, this will be done by first determining each of the density functions in the complete set
of full conditional post-data density functions of the parameters $\theta$, i.e.\ the set of
density functions:
\begin{equation}
\label{equ2}
p(\theta_j\,|\,\theta_{-j},x)\ \ \ \mbox{for $j=1,2,\ldots,k$}
\end{equation}
One of the key features of the approach that will be developed is that it allows any given one of
these density functions to be constructed using whichever one of the three distinct methods of
inference mentioned in the Introduction is regarded as being the most appropriate for the task.

In order to remove a potentially important source of conflict between the three methods of
inference being referred to, the quite natural assumption will be made that during the process of
determining each of the full conditional densities in equation~(\ref{equ2}), the set of
conditioning parameters $\theta_{-j}$ are always treated as being known constants.
This means that usually it will not be permitted that any one of these conditional densities is
determined by first constructing a joint post-data density of the parameter $\theta_j$ and some or
all of the parameters in the set $\theta_{-j}$, and then conditioning this joint density on the
parameters $\theta_{-j}$.
However, making the assumption that has just been made does not generally eliminate the
possibility that the set of full conditional densities in equation~(\ref{equ2}) may be determined
using the methods in question in a way that \pagebreak implies that they are not consistent with
any joint density function of the parameters concerned, i.e.\ these conditional densities may be
incompatible among themselves.
On the other hand, if the full conditional densities under discussion are indeed compatible then,
since, under a mild requirement, a joint density function is uniquely defined by its full
conditional densities, these densities will, in general, define a unique joint post-data density
function for the parameters $\theta$, i.e.\ a unique density $p(\theta\,|\,x)$.

\vspace{3ex}
\noindent
\textbf{Addressing the issue of incompatible full conditional densities}

\vspace{1ex}
\noindent
As discussed in Bowater~(2018a), to check whether full conditional densities of the overall type
being considered are compatible, it may be possible to use a simple analytical method.
In particular, we begin to implement this method by proposing an analytical expression for the
joint density function of the set of parameters $\theta$, then we determine the full conditional
density functions for this joint density, and finally we see whether these conditional densities
are equivalent to the full conditional densities in equation~(\ref{equ2}).
If this equivalence is achieved, then these latter conditional densities clearly must be
compatible. This method has the advantage that generally, in such circumstances, it directly gives
us an analytical expression for the unique joint post-data density $p(\theta\,|\,x)$, \linebreak
i.e.\ under a mild condition, it will be the originally proposed joint density for the
parameters~$\theta$.

By contrast, in situations that will undoubtedly often arise where it is not easy to establish
whether or not the full conditional densities in equation~(\ref{equ2}) are compatible, let us
imagine that we make the pessimistic assumption that they are in fact incompatible.
Nevertheless, even though these full conditional densities could be incompatible, they could be
reasonably assumed to represent the best information that is available for constructing a joint
post-data density function of the parameters $\theta$, or in other words, for constructing the most
suitable density $p(\theta\,|\,x)$.
Therefore, it would seem appropriate to try to find the joint density of the parameters $\theta$
that has full \pagebreak conditional densities that most closely approximate those given in
equation~(\ref{equ2}).

To achieve this objective, let us focus attention on the use of a method that was advocated in a
similar context in Bowater~(2018a), in particular the method that simply consists in making the
assumption that the joint density of the parameters $\theta$ that most closely corresponds to the
full conditional densities in equation~(\ref{equ2}) is equal to the limiting density function of a
Gibbs sampling algorithm (Geman and Geman~1984, Gelfand and Smith~1990) that is based on these
conditional densities with some given fixed or random scanning order of the parameters in question.
Under a fixed scanning order of the model parameters, we will define a single transition of this
type of algorithm as being one that results from randomly drawing a value (only once) from each of
the full conditional densities in equation~(\ref{equ2}) according to some given fixed ordering of
these densities, replacing each time the previous value of the parameter concerned by the value
that is generated.
Let us clarify that it is being assumed that only the set of values for the parameters $\theta$
that are obtained on completing a transition of this kind are recorded as being a newly generated
sample, i.e.\ the intermediate sets of parameter values that are used in the process of making such
a transition do not form part of the output of the algorithm.

To measure how close the full conditional densities of the limiting density function of the general
type of Gibbs sampler being presently considered are to the full conditional densities in
equation~(\ref{equ2}), we can make use of a method that, in relation to its use in a similar
context, was discussed in Bowater~(2018a).
The reasoning that underlies this method can be easily appreciated by first assessing the practical
viability of another specific procedure for verifying the compatibility of the conditional
densities in equation~(\ref{equ2}).

In particular, on the basis of the results in Chen and Ip~(2015), it can be deduced that the
conditional densities in this equation will be compatible if, under a fixed scan\-ning order of the
parameters $\theta$ that is implemented in the way that was just specified, a Gibbs sampling
algorithm based on these full conditional densities satisfies the following \linebreak three
conditions:

\vspace{1.5ex}
\noindent
A) It is positive recurrent for all possible fixed scanning orders. This condition ensures that the
sampling algorithm has at least one stationary distribution for any given fixed scanning order.

\vspace{1.5ex}
\noindent
B) It is irreducible and aperiodic for all possible fixed scanning orders. Together with
condition~A, this condition ensures that the sampling algorithm has a limiting distribu\-tion for
any given fixed scanning order.

\vspace{1.5ex}
\noindent
C) Given conditions~A and~B hold, the limiting density function of the sampling
algo-{\linebreak}rithm needs to be the same over all possible fixed scanning orders.

\vspace{1.5ex}
\noindent
Moreover, when these conditions hold, the joint post-data density function of the parameters
$\theta$ that is directly defined by the full conditional densities in equation~(\ref{equ2}) will
be the unique limiting density function of these parameters referred to in condition~C.
The sufficiency of the conditions~A to~C just listed for establishing the compatibility of any
given set of full conditional densities was proved for a special case in Chen and Ip~(2015), which
is a proof that can be easily extended to the more general case that is currently of interest.

Nevertheless, even if, with respect to the specific type of full conditional densities re\-ferred
to in equation~(\ref{equ2}), we can establish that condition~A and condition~B are satisfied, it
will usually be impossible, in practice, to determine whether condition~C is satisfied.
From an alternative perspective, if we assume that the full conditional densities in this equation
are in fact incompatible, then if conditions~A and~B are satisfied, it would ap\-pear to be useful
(\hspace{0.1em}with reference to condition~C\hspace{0.1em}) to analyse how the limiting density
function of a Gibbs sampler based on these full conditional densities varies over a reasonable
number of very distinct fixed scanning orders of the sampler.
If within such an analysis, the variation of this limiting density with respect to the scanning
order of the parameters $\theta$ can be classified as small, negligible or undetectable, then this
should give us reassurance that the full conditional densities in equation~(\ref{equ2}) are,
respectively according to such classifications, close, very close or at least very close, to the
full conditional densities of the limiting density of a Gibbs sampler of the type that is of main
interest, i.e.\ a Gibbs sampler that is based on any given fixed or random scanning order of the
\linebreak parameters concerned.

In trying to choose the scanning order of this type of Gibbs sampler such that it has a limiting
density function that corresponds to a set of full conditional densities that most accurately
approximate the density functions in equation~(\ref{equ2}), a good general choice would arguably
be, what will be referred to as, a uniform random scanning order.
Under this type of scanning order, a transition of the Gibbs sampling algorithm in question will be
defined as being one that results from generating a value from one of the full conditional
densities in equation~(\ref{equ2}) that is chosen at random, with the same probability of $1/k$
being given to any one of these densities being selected, and then treating the generated value as
the updated value of the parameter concerned.

However, it can be easily shown that independent of whether or not the set of full conditional
densities in equation~(\ref{equ2}) are compatible, the last full conditional density in this set
that is sampled from in completing a given fixed scanning order will be one of the full conditional
densities of the limiting density function of the type of Gibbs sampler being discussed that uses
such a fixed scanning order.
This therefore provides a reason for perhaps deciding, in certain applications, that the limiting
density of a Gibbs sampler of the general type in question most satisfactorily corresponds to the
full conditional densities in equation~(\ref{equ2}) when a given fixed rather than a uniform random
scanning order of the parameters $\theta$ is used.

\pagebreak
\noindent
\textbf{Conventional simulation issues}

\vspace{1ex}
\noindent
As with all Gibbs samplers it is important to verify in implementing strategies of the type just
mentioned that the sampler concerned has converged to its limiting density function within the
restricted number of transitions of the sampler that can be observed in practice.
To do this, we can make use of standard methods for analysing the convergence of Monte Carlo Markov
chains described in, for example, Gelman and Rubin~(1992), Cowles and Carlin~(1996) and Brooks and
Roberts~(1998).
However, the use of such convergence diagnostics may be considered to be slightly more important in
the case of present interest in which the full conditional densities on which the Gibbs sampler is
based could be incompatible, since, compared to the case where these densities are known to be
compatible, there is likely to be, in practice, a little more concern that the Gibbs sampler may
not actually have a limiting density function, even though in reality the genuine risk of this may
still be extremely low.

A notable advantage of the general method for finding a suitable joint post-data density for the
parameters $\theta$ that has just been outlined is that it can directly achieve what is often the
main goal of a standard application of the Gibbs sampler, namely that of obtaining good
approximations to the expected values of functions of the parameters of a model over the post-data
or posterior density for these parameters that is of interest, i.e.\ expected values of the
following type:
\vspace{2ex}
\[
\mbox{E}[h(\theta)\,|\,x] = \int_{\mbox{\footnotesize{$\mathbb{R}^k$}}} h(\theta)p(\theta\,|\,x)
d\theta
\vspace{2ex}
\]
where $p(\theta\,|\,x)$ is a given post-data density function of the parameters $\theta$, while
$h(\theta)$ is \linebreak any given function of these parameters.
To be more specific, this kind of expected value may, of course, be approximated using the Monte
Carlo estimator:
\pagebreak
\[
\frac{1}{N-b}\hspace{0.05em} \sum_{i\hspace{0.05em}=\hspace{0.05em}b\hspace{0.05em}+1}^{N}
h(\theta_1^{(i)},\theta_2^{(i)},\ldots,\theta_k^{(i)})
\vspace{1.5ex}
\]
where $\theta_1^{(i)}, \theta_2^{(i)},\ldots, \theta_k^{(i)}$ is the $i$th sample of parameter
values among the $N$ samples generated by the sampler in total, and $b$ is the number of initial
samples that are classified as belonging to the burn-in phase of the sampler.

\vspace{3ex}
\subsection{Bayesian inference}
\label{sec3}

As was in effect done so by Bayes in his famous paper Bayes~(1763), it will be assumed that
Bayesian inference depends on three key concepts. First, Bayes' theorem as a purely mathematical
expression. Second, the justification of the application of this theorem to well-understood
physical experiments, e.g.\ random spins of a wheel or random draws of a ball from an urn of balls.
Finally, something which will be referred to as Bayes' analogy, which is the type of analogy that
can be made between the uncertainty that surrounds the outcomes of the kind of physical experiments
just mentioned to which Bayes' theorem can be very naturally applied, and the uncertainty that
surrounds what are the true values of any unknown real-world quantities that are of interest.

By using this latter concept, we can justify the use of Bayesian inference in a much wider range of
applications than is allowed by only using the first two concepts.
However, depending on the type of application, the Bayes' analogy may be a good analogy or a poor
analogy, which is something that needs to be taken into account when assessing the adequacy of any
given application of the Bayesian method.

In keeping with the notation defined in the Introduction, the post-data or posterior density
function of the parameter $\theta_j$ given all other model parameters $\theta_{-j}$ can be
expressed according to Bayes' theorem as follows:
\[
p(\theta_j\,|\,\theta_{-j},x) = \mathtt{C}_0\hspace{0.05em} g(x\,|\,\theta)
p(\theta_j\,|\,\theta_{-j})
\]
where $p(\theta_j\,|\,\theta_{-j})$ is the pre-data or prior density function of the parameter
$\theta_j$ given the parameters $\theta_{-j}$, while $\mathtt{C}_0$ is a normalising constant.

In this paper, we will exclude from consideration two methods of inference that are often referred
to as `objective' forms of Bayesian inference. The first of these methods consists in always
specifying the prior density $p(\theta_j\,|\,\theta_{-j})$ as being a uniform or flat den\-sity
function over all values of $\theta_j$.
This implies, though, that the Bayes' analogy must be broken due to this prior density being
improper and/or due to the posterior density of any given population quantity of interest
$h(\theta_j)$ conditional on the parameters $\theta_{-j}$ possessing, in general, the property of
being dependent on the parameterisation of the sampling model, which of course is a very
undesirable property for this posterior density to have.
On the other hand, the second type of method entails specifying the prior density
$p(\theta_j\,|\,\theta_{-j})$ such that it depends on the sampling model, i.e.\ allowing what is
known about the parameter $\theta_j$ to depend on how we intend to collect more information about
this parameter, however doing this clearly again breaks the Bayes' analogy.
A famous example of a type of prior density that is specified in this way is a prior density that
is derived by applying Jeffreys' rule, see Jeffreys~(1961), although many other prior densities of
this kind have been proposed, see for example, Kass and Wasserman~(1996).

To conclude, it can be strongly argued that, due to the Bayes' analogy being clearly broken, the
application of either of the two methods of inference that have just been mentioned should not
really be regarded as being an application of the Bayesian approach to inference at all.

\vspace{3ex}
\subsection{Organic fiducial inference}
\label{sec4}

We will now outline some of the key concepts that underlie the theory of organic fiducial
inference. Descriptions of other important concepts on which this theory is based, along with
further details about the concepts that will be outlined here and about the overall theory itself,
can be found in Bowater~(2019a). Throughout this section, it will be assumed that the values of the
parameters in the set $\theta_{-j}$ are known.

\vspace{3ex}
\noindent
{\bf Fiducial statistics}

\vspace{1ex}
\noindent
A fiducial statistic $Q(x)$ will be defined as being a univariate statistic of the sample $x$ that
can be regarded as efficiently summarising the information that is contained in this sample about
the only unknown parameter $\theta_j$, given the values of other statistics that do not provide any
information about this parameter, i.e.\ ancillary statistics. If, in any given case, there exists a
univariate sufficient statistic for $\theta_j$, then this would naturally be chosen to be the
fiducial statistic for that case. In other cases, it may well make good sense to choose this
statistic $Q(x)$ to be the maximum likelihood estimator of $\theta_j$.

For ease of presentation, we will assume, in what follows, that the choice of the fiducial
statistic can be justified without reference to any particular ancillary statistics.

\vspace{3ex}
\noindent
{\bf Data generating algorithm}

\vspace{1ex}
\noindent
Independent of the way in which the data set $x$ was actually generated, it will be assumed that
this data set was generated by the following algorithm:

\vspace{2ex}
\noindent
1) Generate a value $\gamma$ for a continuous one-dimensional random variable $\Gamma$, which has a
density function $\pi_0(\gamma)$ that does not depend on the parameter $\theta_j$.

\vspace{2ex}
\noindent
2) Determine a value $q(x)$ for the fiducial statistic $Q(x)$ by setting $\Gamma$ equal to $\gamma$
and $Q(x)$ equal to $q(x)$ in the following expression for the statistic $Q(x)$, which effectively
should define the way in which this statistic is distributed:
\begin{equation}
\label{equ3}
Q(x)=\varphi(\Gamma, \theta_j)
\end{equation}
where the function $\varphi(\Gamma, \theta_j)$ is specified so that it satisfies the following
conditions:

\pagebreak
\noindent
a) The density or mass function of $Q(x)$ that is, in effect, defined by equation~(\ref{equ3}) is
\linebreak equal to what it would have been if $Q(x)$ had been determined on the basis of the data
set $x$.\\
b) The only random variable upon which $\varphi(\Gamma, \theta_j)$ depends is the variable
$\Gamma$.

\vspace{2ex}
\noindent
3) Generate the data set $x$ from its sampling density or mass function
$g(x\,|\,\theta_1,\theta_2,\ldots,\theta_k)$ conditioned on the statistic $Q(x)$ being equal to its
already generated value $q(x)$.

\vspace{3ex}
In the context of this algorithm, the variable $\Gamma$ is referred to as the primary random
variable (primary r.v.).

\vspace{3ex}
\noindent
{\bf Strong fiducial argument}

\vspace{1ex}
\noindent
This is the argument that the density function of the primary r.v.\ $\Gamma$ after the data have
been observed, i.e.\ the post-data density function of $\Gamma$, should be equal to the pre-data
density function of $\Gamma$, i.e.\ the density function $\pi_0(\gamma)$ as defined in step~1 of
the data generating algorithm just presented.

\vspace{3ex}
\noindent
{\bf Moderate fiducial argument}

\vspace{1ex}
\noindent
It will be assumed that this argument is only applicable if, on observing the data $x$, there
exists some positive measure set of values of the primary r.v.\ $\Gamma$ over which the pre-data
density function $\pi_0(\gamma)$ was positive, but over which the post-data density function of
$\Gamma$, which will be denoted as the density function $\pi_1(\gamma)$, is necessarily zero.
Under this condition, it is the argument that, over the set of values of $\Gamma$ for which the
density function $\pi_1(\gamma)$ is necessarily positive, the relative height of this function
should be equal to the relative height of the density function $\pi_0(\gamma)$, i.e.\ the heights
of these two functions should be proportional.

\pagebreak
\noindent
{\bf Weak fiducial argument}

\vspace{1ex}
\noindent
This argument will be assumed to be only applicable if neither the strong nor the moderate fiducial
argument is considered to be appropriate.
It is the argument that, over the set of values of the primary r.v.\ $\Gamma$ for which the
post-data density function $\pi_1(\gamma)$ is necessarily positive, the relative height of this
function should be equal to the relative height of the pre-data density function $\pi_0(\gamma)$
multiplied by weights on the values of $\Gamma$ determined by a given function over the parameter
$\theta_j$ that was specified before the data were observed. This latter function is called the
global pre-data function of $\theta_j$. Let us now define this function.

\vspace{3ex}
\noindent
{\bf Global pre-data (GPD) function}

\vspace{1ex}
\noindent
The global pre-data (GPD) function $\omega_G(\theta_j)$ is used to express pre-data knowledge, or a
lack of such knowledge, about the only unknown parameter $\theta_j$.
This function may be any given non-negative and upper bounded function of the parameter $\theta_j$.
It is a function that only needs to be specified up to a proportionality constant, in the sense
that, if it is multiplied by a positive constant, then the value of the constant is redundant.
Unlike a Bayesian prior density, it is not controversial to use a GPD function that is not globally
integrable.

\vspace{3ex}
\noindent
{\bf A principle for defining the fiducial density $f(\theta_j\,|\,\theta_{-j},x)$}

\vspace{1ex}
\noindent
Let us now consider a principle for defining the post-data density of $\theta_j$ conditional on
\linebreak the parameters $\theta_{-j}$, which given that it will be derived using a type of
fiducial inference, will be called the fiducial density of $\theta_j$ conditional on $\theta_{-j}$,
and will be denoted as the density $f(\theta_j\,|\,\theta_{-j},x)$. To be able to use this
principle, the following condition must be satisfied.

\pagebreak
\noindent
{\bf Condition 1}

\vspace{1ex}
\noindent
Let $G_x$ and $H_x$ be, respectively, the sets of all the values of the primary r.v.\ $\Gamma$ and
the parameter $\theta_j$ for which the density functions of these variables must necessarily be
positive in light of having observed only the value of the fiducial statistic $Q(x)$, i.e.\ the
value $q(x)$, and not any other information in the data set $x$.
To clarify, any set of values of $\Gamma$ or any set of values of $\theta_j$ that are regarded as
being impossible after the statistic $Q(x)$ has been observed can not be contained in the set $G_x$
or the set $H_x$ respectively.
Given this notation, the present condition will be satisfied if, on substituting the variable
$Q(x)$ in equation~(\ref{equ3}) by its observed value $q(x)$, this equation would define a
bijective mapping between the set $G_x$ and the set $H_x$.

\vspace{2ex}
Under this condition, the full conditional fiducial density $f(\theta_j\,|\,\theta_{-j},x)$ is
defined by setting $Q(x)$ equal to its observed value $q(x)$ in equation~(\ref{equ3}), and then
treating the \linebreak value $\theta_j$ in this equation as being a realisation of the random
variable $\Theta_j$, to give the expression:
\vspace{0.75ex}
\begin{equation}
\label{equ12}
q(x)=\varphi(\Gamma, \Theta_j)
\vspace{1.25ex}
\end{equation}
except that, instead of the variable $\Gamma$ necessarily having the density function
$\pi_0(\gamma)$ as defined in step~1 of the data generating algorithm, it will be assumed to have
the post-data density function of this variable as defined by:
\vspace{2ex}
\[
\pi_1(\gamma) = \left\{
\begin{array}{ll}
\mathtt{C}_1\hspace{0.05em} \omega_G(\theta_j(\gamma))\hspace{0.05em} \pi_0(\gamma) \ & \mbox{if
$\gamma \in G_x$}\\[1ex]
0 & \mbox{otherwise}
\end{array}
\right.
\vspace{2ex}
\]
where $\theta_j(\gamma)$ is the value of the variable $\Theta_j$ that maps on to the value $\gamma$
of the variable $\Gamma$ according to equation~(\ref{equ12}), the function
$\omega_G(\theta_j(\gamma))$ is the GPD function of $\theta_j$ defined earlier, and $\mathtt{C}_1$
is a normalising constant.

\pagebreak
Notice that if, on substituting the variable $Q(x)$ by the value $q(x)$, equation~(\ref{equ3})
de\-fines an injective mapping from the set of values $\{\gamma: \pi_0(\gamma) >0\}$ for the
variable $\Gamma$ to the space of the parameter $\theta_j$, then the GPD function
$\omega_G(\theta_j)$ expresses in effect our pre-data beliefs about $\theta_j$ relative to what is
implied by using the strong fiducial argu-{\linebreak}ment.
By doing so, it determines whether the strong, moderate or weak fiducial argument is used to make
inferences about $\theta_j$, and also the way in which the latter two arguments influence the
inferential process.

In the case where nothing or very little was known about the parameter $\theta_j$ before the data
were observed, it would generally seem reasonable to choose the GPD function of the parameter
$\theta_j$ to be equal to a positive constant over the entire space of this parameter. Under the
assumption that there exists an injective mapping from the space of\hspace{0.03em} $\Gamma$ to the
space of $\theta_j$ of the type just mentioned, choosing the GPD function $\omega_G(\theta_j)$ in
this way implies that the post-data density $\pi_1(\gamma)$ will be equal to the pre-data density
$\pi_0(\gamma)$, i.e.\ inferences will be made about $\theta_j$ by using the strong fiducial
argument.
The use of the theory of fiducial inference being presently considered in this special case is
discussed to some extent in Bowater~(2019a), but more extensively in Bowater~(2018a), where in fact
a specific version of organic fiducial inference is applied to examples of this particular nature
that is referred to as subjective fiducial inference.

\vspace{3ex}
\noindent
{\bf Other ways of defining the fiducial density $f(\theta_j\,|\,\theta_{-j},x)$}

\vspace{1ex}
\noindent
In cases where the principle just described can not be applied, i.e.\ when Condition~1 does not
hold, we may well be able to define the fiducial density $f(\theta_j\,|\,\theta_{-j},x)$ using the
alternative principle for this purpose that was presented in Section~3.4 of Bowater~(2019a) as
Principle~2, or it may well be considered acceptable to define this fiducial density using the kind
of variations on this latter principle that were discussed in Sections~7.2 and~8 of this earlier
paper.
The alternative principle in question, which is \pagebreak particularly useful in cases where the
data are discrete or categorical, relies on the concept of a local pre-data (LPD) function for
expressing additional information concerning the pre-data beliefs that were held about the
parameter $\theta_j$ to that which is expressed by the GPD function for $\theta_j$. The concept of
a LPD function is also detailed in Bowater~(2019a).

\vspace{3ex}
\subsection{Bispatial inference}
\label{sec5}

The type of bispatial inference that will be incorporated into the theory being developed in the
present paper will be the special form of bispatial inference that was laid out in Section~3 of
Bowater~(2019b). Let us now outline the key concepts on which this type of bispatial inference is
based. Further details about these concepts and a broader discussion of the specific method of
inference in question can be found in Bowater~(2019b). As in the previous section, the values of
the parameters in the set $\theta_{-j}$ will be assumed to be known.

\vspace{3ex}
\noindent
{\bf Scenario of interest}

\vspace{1ex}
\noindent
This scenario is characterised by there having been a substantial degree of belief before the data
were observed that the only unknown parameter $\theta_j$ lay in a narrow interval $[\theta_{j0},
\theta_{j1}]$, but if, on the other hand, $\theta_j$ had been conditioned not to lie in this
interval, then there would have been no or very little pre-data knowledge about $\theta_j$ over all
of its allowable values outside of the interval in question.
Among the three common types of pre-data opinion we may hold about the parameter $\theta_j$ that
were highlighted in the Introduction, this scenario is clearly consistent with holding the second
type of opinion.

\vspace{3ex}
\noindent
{\bf Test statistics}

\vspace{1ex}
\noindent
In the context of bispatial inference, a test statistic $T(x)$, which will also be denoted simply
by the value $t$, is specified such that it satisfies two criteria.
First, this statistic must fit within the broad definition of a fiducial statistic that was given
in the previous section.
Therefore, this could mean that a particular choice of the statistic $T(x)$ can only be justified
with reference to given ancillary statistics, however, similar to how we proceeded in the previous
section, we will assume here, for ease of presentation, that this is not the case.

The second criterion is that if $F(t \,|\, \theta_j)$ is the cumulative distribution function of
the unobserved test statistic $T(X)$ evaluated at its observed value $t$ given a value for the
parameter $\theta_j$, i.e.\ $F(t \,|\, \theta_j) = P(T(X) \leq t\,|\,\theta_j)$, and if
$\Fprime(t \,|\, \theta_j)$ is equal to the probability $P(T(X) \geq t\,|\,\theta_j)$, then it is
necessary that, over the set of allowable values for $\theta_j$, the probabilities
$F(t \,|\, \theta_j)$ and $1 - \Fprime(t \,|\, \theta_j)$ strictly decrease as $\theta_j$
increases.

\vspace{3ex}
\noindent
{\bf Parameter and sampling space hypotheses}

\vspace{1ex}
\noindent
Under this definition of a test statistic $T(x)$, if the condition:
\begin{equation}
\label{equ4}
F(t \,|\, \theta_j=\theta_{j0}) \leq \Fprime(t \,|\, \theta_j=\theta_{j1})
\end{equation}
holds, where the values $\theta_{j0}$ and $\theta_{j1}$ are as defined at the start of this
section, then the parameter space hypothesis $H_{P}$ and the sampling space hypothesis $H_{S}$ will
be defined \nolinebreak as:
\begin{gather}
H_{P}: \theta_j \geq \theta_{j0} \label{equ5}\\
H_{S}: \rho( T(X^*) \leq t) \leq F(t\,|\,\theta_j = \theta_{j0}) \label{equ6}
\end{gather}
where $X^*$ is an as-yet-unobserved second sample of values drawn from the sampling density of
interest, i.e.\ the density $g(x\,|\,\theta)$, that is the same size as the observed (first) sample
$x$, i.e.\ it consists of $n$ observations, and where $\rho(A)$ is the unknown population
proportion of times that condition $A$ is satisfied.
On the other hand, if the condition in equation~(\ref{equ4}) does not hold, then the hypotheses in
question \pagebreak will be defined as:
\begin{gather}
H_{P}: \theta_j \leq \theta_{j1} \label{equ10}\\
H_{S}: \rho( T(X^*) \geq t) \leq \Fprime(t\,|\,\theta_j = \theta_{j1}) \label{equ7}
\end{gather}

\vspace{0.5ex}
Given the way that the test statistic $T(x)$ was just defined, it can be easily appreci\-ated that
the hypotheses $H_{P}$ and $H_{S}$ in equations~(\ref{equ5}) and~(\ref{equ6}) are equivalent, and
also that these hypotheses as defined in equations~(\ref{equ10}) and~(\ref{equ7}) are equivalent.
In addition, observe that the probabilities $F(t\,|\,\theta_j = \theta_{j0})$ and
$\Fprime(t\,|\,\theta_j = \theta_{j1})$ that appear in the definitions of the hypotheses $H_{S}$ in
equations~(\ref{equ6}) and~(\ref{equ7}) would be the standard one-sided P values that would be
calculated on the basis of the data set $x$ if the null hypotheses were regarded as being the
hypotheses $H_{P}$ that correspond to the two hypotheses \linebreak $H_{S}$ in question.

\vspace{3ex}
\noindent
{\bf Inferential process}

\vspace{1ex}
\noindent
It will be assumed that inferences are made about the parameter $\theta_j$ by means of the
following three-step process:

\vspace{2ex}
\noindent
Step 1: Assessment of the likeliness of the hypothesis $H_{P}$ being true using only pre-data
knowledge about the parameter $\theta_j$, with special attention being given to evaluating the
likeliness of the hypothesis that $\theta_j$ lies in the interval $[\theta_{j0}, \theta_{j1}]$,
which is an  hypothesis that is always included in the hypothesis $H_{P}$.
It is not necessary that this assessment is expressed in terms of a formal measure of uncertainty,
e.g.\ a probability does not need to be assigned to the hypothesis $H_{P}$.

\vspace{2ex}
\noindent
Step 2: Assessment of the likeliness of the hypothesis $H_{S}$ being true after the data $x$ have
been observed, leading to the assignment of a probability to this hypothesis, which will be denoted
as the probability $\kappa$.
In carrying out this assessment, all relevant factors ought to be taken into account including, in
particular:
(a) the size of \pagebreak the one-sided P value that appears in the definition of the hypothesis
$H_{S}$, i.e.\ the value $F(t\,|\,\theta_j = \theta_{j0})$ or the value
$\Fprime(t\,|\,\theta_j = \theta_{j1})$, (b) the assessment made in Step~1, and (c) the known
equivalency between the hypotheses $H_{P}$ and $H_{S}$.

\vspace{2ex}
\noindent
Step 3: Conclusion about the probability of the hypothesis $H_{P}$ being true having taken into
account the data $x$. This is directly implied by the assessment made in Step~2 due to the
equivalence of the hypotheses $H_{P}$ and $H_{S}$.

\vspace{3ex}
\noindent
{\bf In combination with organic fiducial inference}

\vspace{1ex}
\noindent
It was described in Bowater~(2019b) how the type of bispatial inference under discussion can be
extended from allowing us to simply determine a post-data probability for the hypothesis $H_{P}$
being true, i.e.\ the probability $\kappa$, to allowing us to determine an entire post-data density
function for the parameter $\theta_j$. As was the case in this earlier paper, we will again favour
doing this in an indirect way by combining bispatial inference as has just been detailed with
organic fiducial inference as was summarised in Section~\ref{sec4}. In particular, the method that
we will choose to adopt to achieve the goal in question will be essentially the method that was put
forward in Section~4.2 of Bowater~(2019b). Let us now give briefly outline this method.

To begin with, in applying the method concerned, we assume that both the post-data density function
of $\theta_j$ conditional on $\theta_j$ lying in the interval $[\theta_{j0}, \theta_{j1}]$, and the
post-data density function of $\theta_j$ conditional on $\theta_j$ not lying in this interval are
derived under the paradigm of organic fiducial inference, i.e.\ they are fiducial density
functions, and let us therefore denote these density functions by
$f(\theta_j\,|\, \theta_j \in [\theta_{j0}, \theta_{j1}], x)$ and
$f(\theta_j\,|\, \theta_j \notin [\theta_{j0}, \theta_{j1}], x)$ respectively.
Since it has been assumed that, under the condition that $\theta_j$ does not lie in the interval
$[\theta_{j0}, \theta_{j1}]$, nothing or very little would have been \linebreak known about
$\theta_j$ before the data were observed, it would seem quite natural, in deriving the latter of
these fiducial densities $f(\theta_j\,|\, \theta_j \notin [\theta_{j0}, \theta_{j1}], x)$, to use a
GPD function for $\theta_j$ that has the following form:
\vspace{1.5ex}
\[
\omega_G (\theta_j) = \left\{
\begin{array}{ll}
0\ & \mbox{if $\theta_j \in [\theta_{j0}, \theta_{j1}]$}\\[1.25ex]
a & \mbox{otherwise}
\end{array}
\right.
\vspace{2ex}
\]
where $a>0$, which would be classed as a neutral GPD function using the terminology of
Bowater~(2019a).

On the basis of this GPD function, the fiducial density
$f(\theta_j\,|\, \theta_j \notin [\theta_{j0}, \theta_{j1}], x)$ can of\-ten be derived by applying
the moderate fiducial argument under the principle that was outlined in Section~\ref{sec4}, i.e.\
Principle~1 of Bowater~(2019a). Alternatively, in accordance with what was also advocated in
Bowater~(2019a), this fiducial density can be more generally defined, with respect to the same GPD
function for $\theta_j$, by the following ex-{\linebreak}pression:
\vspace{1ex}
\begin{equation}
\label{equ8}
f(\theta_j\,|\, \theta_j \notin [\theta_{j0}, \theta_{j1}], x) = \mathtt{C}_2\hspace{0.05em}
f_{S}(\theta_j\,|\, x)
\vspace{1.5ex}
\end{equation}
where $\mathtt{C}_2$ is a normalising constant, and $f_{S}(\theta_j\,|\,x)$ is a fiducial density
for $\theta_j$ derived us-{\linebreak}ing either Principle~1 or Principle~2 of Bowater~(2019a) that
would be regarded as being a suitable fiducial density for $\theta_j$ in a general scenario where
it is assumed that there was no or very little pre-data knowledge about $\theta_j$ over all
possible values of $\theta_j$.

To construct the fiducial density of $\theta_j$ conditional on $\theta_j$ lying in the interval
$[\theta_{j0}, \theta_{j1}]$, i.e.\ the density $f(\theta_j\,|\, \theta_j\hspace{-0.1em}
\in\hspace{-0.05em} [\theta_{j0}, \theta_{j1}], x)$, the method being considered relies on quite a
general type of GPD function for $\theta_j$. In particular, it is assumed that this GPD function
has the following form:
\vspace{2ex}
\begin{equation}
\label{equ9}
\oline{\omega}_G (\theta_j) = \left\{
\begin{array}{ll}
1 + \nu h(\theta_j)\ & \mbox{if $\theta_j \in [\theta_{j0}, \theta_{j1}]$}\\[1.25ex]
0 & \mbox{otherwise}
\end{array}
\right.
\vspace{2.5ex}
\end{equation}
where $\nu \geq 0$ is a given constant and $h(\theta_j)$ is a continuous unimodal \pagebreak
density function on the interval $[\theta_{j0}, \theta_{j1}]$ that is equal to zero at the limits
of this interval. On the basis of this GPD function, the fiducial density
$f(\theta_j\,|\, \theta_j \in [\theta_{j0}, \theta_{j1}], x)$ can often be derived by again using
the principle detailed in Section~\ref{sec4} (i.e.\ Principle~1 of Bowater~2019a), but this time by
calling upon the weak fiducial argument.
Alternatively, in accordance with what was also advocated in Bowater~(2019a), this fiducial density
can be more generally defined, with respect to the same GPD function for $\theta_j$, in the
following way:
\vspace{0.5ex}
\begin{equation}
\label{equ19}
f(\theta_j\,|\,\theta_j \in [\theta_{j0}, \theta_{j1}], x)
= \mathtt{C}_3\hspace{0.15em} \oline{\omega}_G (\theta_j) f_{S}(\theta_j\,|\,x)
\vspace{0.5ex}
\end{equation}
where the fiducial density $f_{S}(\theta_j\,|\,x)$ is specified as it was immediately after
equation~(\ref{equ8}), and $\mathtt{C}_3$ is a normalising constant.

Now, if in using the method of bispatial inference outlined immediately before the current
discussion, the hypothesis $H_{P}$, i.e.\ the hypothesis in equation~(\ref{equ5}) or
equation~(\ref{equ10}), is assigned a sensible post-data probability $\kappa$, i.e.\ a probability
above a very low limit that is defined in Bowater~(2019b), then given the two conditional post-data
densities for $\theta_j$ that have just been specified, i.e.\ the fiducial densities
$f(\theta_j\,|\,\theta_j \in [\theta_{j0}, \theta_{j1}], x)$ and
$f(\theta_j\,|\, \theta_j \notin [\theta_{j0}, \theta_{j1}], x)$, we have sufficient information to
determine a valid post-data density function of $\theta_j$ over all values of $\theta_j$.
Hopefully, it is fairly clear why this is the case, nevertheless the reader is referred to
Bowater~(2019b) for a more detailed account of the derivation of this latter post-data density
function.
In the rest of this paper, we will denote this overall post-data density function of $\theta_j$ as
the density $b(\theta_j\,|\, \theta_{-j}, x)$ to indicate that it was derived using bispatial
inference.

However, there is an important final issue that needs to be resolved, which is how the value of the
constant $\nu$ in equation~(\ref{equ9}) is chosen. Using the method being discussed, this constant
must in fact be chosen such that the overall post-data density
$b(\theta_j\,|\, \theta_{-j}, x)$ is made equivalent to a fiducial density function for $\theta_j$
that is based on a continuous GPD function for $\theta_j$ over all values of $\theta_j$, but except
for the \pagebreak way in which this GPD function is specified, is based on the same assumptions as
were used to derive the fiducial density $f_{S}(\theta_j\,|\,x)$. In general, a value for $\nu$
will exist that satisfies this condition and it will be a unique value.
Placing this condition on the choice of $\nu$ can be viewed as not restricting excessively the way
we are allowed to express our pre-data knowledge about the parameter $\theta_j$, while it ensures
that the density function $b(\theta_j\,|\, \theta_{j}, x)$ possesses, in general, the usually
desirable property of being continuous over all values of $\theta_j$.

\vspace{3ex}
\noindent
{\bf Post-data opinion (PDO) curve}

\vspace{1ex}
\noindent
Observe that in using the method of inference that has just been outlined, the assessment of the
likeliness of the hypothesis $H_S$ in either equation~(\ref{equ6}) or equation~(\ref{equ7}) will,
in gen\-eral, depend on the values of the parameters in the set $\theta_{-j}$.
This of course will be partially due to the effect that the values of these parameters can have on
the one-sided P value that appears in the definition of this hypothesis, i.e.\ their effect on the
value \linebreak $F(t\,|\,\theta_j = \theta_{j0})$ or the value
$\Fprime(t\,|\,\theta_j = \theta_{j1})$.
As a result, to implement the method of inference under discussion within the overall framework for
determining a joint post-data density of all the model parameters $\theta$ that was put forward in
Section~\ref{sec2}, we will generally wish to assign not just one, but various probabilities to the
hypothesis $H_{S}$ conditional on the values of the parameters $\theta_{-j}$.

It is possible though to simplify matters greatly by assuming that the probability that is assigned
to any given hypothesis $H_{S}$, and to also therefore its corresponding hypothesis $H_{P}$, i.e.\
the probability $\kappa$, will be the same for any fixed value of the one-sided P value that
appears in the definition of the hypothesis $H_{S}$ no matter what values are actually taken by the
parameters in the set $\theta_{-j}$.
By making this assumption, which is arguably a reasonable assumption in many practical situations,
the probability $\kappa$ be\-comes a mathematical function of the one-sided P value that appears in
the definition of the hypothesis $H_{S}$ concerned.
As was the case in Bowater~(2019b), this function \pagebreak will be called the post-data opinion
(PDO) curve for the parameter $\theta_j$ conditional on the \linebreak parameters~$\theta_{-j}$.

\vspace{3ex}
\section{Examples}

We will now present various examples of the application of the overall theory that was outlined in
previous sections, i.e.\ the theory of integrated organic inference.

\vspace{2ex}
\subsection{Inference about a univariate normal distribution}
\label{sec10}
\label{sec6}

Let us begin by considering what can be referred to as Student's problem, that is, the standard
problem of making inferences about the mean $\mu$ of a normal density function, when its variance
$\sigma^2$ is unknown, on the basis of a sample $x$ of size $n$, i.e.\ $x=\{x_1,x_2,\ldots,x_n\}$,
drawn from the density function concerned.

If $\sigma^2$ was known, a sufficient statistic for $\mu$ would be the sample mean $\bar{x}$, which
there\-fore, in applying the theory of fiducial inference outlined in Section~\ref{sec4}, can
naturally be assumed to be the fiducial statistic $Q(x)$ in this particular case. Based on this
assump-{\linebreak}tion and given a value for $\sigma^2$, equation~(\ref{equ3}) can be expressed
as:
\vspace{0.5ex}
\begin{equation}
\label{equ28}
\bar{x}=\varphi(\Gamma,\mu)=\mu+(\sigma/\sqrt{n}\hspace{0.1em})\hspace{0.05em}\Gamma
\vspace{0.5ex}
\end{equation}
where the primary r.v.\ $\Gamma \sim \mbox{N}(0,1)$.
If nothing or very little was known about $\mu$ before the data $x$ were observed, then it would be
quite natural to specify the GPD function for $\mu$ as follows: $\omega_G(\mu)=a$ for
$\mu \in (-\infty,\infty)$, where $a>0$, which is indeed in keeping with how this function would be
chosen using a criterion mentioned in Section~\ref{sec4}.
Using the principle outlined in this earlier section for deriving the fiducial density
$f(\theta_j\,|\,\theta_{-j},x)$, and in particular using equation~(\ref{equ12}), this would imply
that the fiducial \pagebreak density of $\mu$ given $\sigma^2$, i.e.\ the density
$f(\mu\,|\,\sigma^2,x)$, is defined by:
\begin{equation}
\label{equ11}
\mu\, |\, \sigma^2, x \sim \mbox{N} (\bar{x}, \sigma^2/n)
\end{equation}

On the other hand, if $\mu$ was known, a sufficient statistic for $\sigma^2$ would be
$\bm\hat{\sigma}^2 = (1/n) \sum_{i=1}^{n} (x_i-\mu)^2$, which therefore, in applying again the
theory of Section~\ref{sec4}, will be assumed to be the statistic $Q(x)$ in this case.
Based on this assumption and given a value for $\mu$, equation~(\ref{equ3}) can be expressed as:
\begin{equation}
\label{equ29}
\bm\hat{\sigma}^2 = \varphi(\Gamma,\sigma^2)=(\sigma^2/n)\Gamma
\end{equation}
where the primary r.v.\ $\Gamma$ has a $\chi^2$ distribution with $n$ degrees of freedom.
If there was no or very little pre-data knowledge about $\sigma^2$, it would be quite natural to
specify the GPD function for $\sigma^2$ as follows:
\begin{equation}
\label{equ30}
\omega_G(\sigma^2) = b\ \ \ \, \mbox{if $\sigma^2 \geq 0$ and zero otherwise}
\end{equation}
where $b>0$. Again using the principle detailed in Section~\ref{sec4} for deriving the fiducial
density $f(\theta_j\,|\,\theta_{-j},x)$, this would imply that the fiducial density
\vspace{0.5ex} $f(\sigma^2\,|\,\mu,x)$ is defined by:
\begin{equation}
\label{equ17}
\sigma^2\,|\,\mu,x \sim \mbox{Inv-Gamma}\, (\alpha = n/2,\hspace{0.1em}
\beta = n\bm\hat{\sigma}^2/2)
\end{equation}
i.e.\ it is an inverse gamma density function with shape parameter $\alpha$ equal to $n/2$ and
scale parameter $\beta$ equal to $n\bm\hat{\sigma}^2/2$.

It can be shown that the full conditional fiducial densities $f(\mu\,|\,\sigma^2,x)$ and
$f(\sigma^2\,|\,\mu,x)$ as they have just been specified are compatible and the joint density
function of $\mu$ and $\sigma^2$ that they directly define is unique.
This density function is therefore the joint fiducial density of $\mu$ and $\sigma^2$.
In particular, the marginal density of $\mu$ over this joint fiducial density is given by:
\pagebreak
\begin{equation}
\label{equ16}
\mu\,|\,x \sim \mbox{Non-standardised}\ t_{n-1}(\bar{x}, s/\sqrt{n}\hspace{0.1em})
\vspace{1ex}
\end{equation}
where $s$ is the sample standard deviation, i.e.\ it is a non-standardised Student $t$ density
function with $n-1$ degrees of freedom, location parameter equal to $\bar{x}$ and scaling
pa\-rameter equal to $s/\sqrt{n}$ (which are settings that of course make it a very familiar
mem-{\linebreak}ber of this particular family of density functions), while the marginal density of
$\sigma^2$ over \linebreak the joint fiducial density of $\mu$ and $\sigma^2$ in question is
given by:
\begin{equation}
\label{equ15}
\sigma^2\,|\,x \sim \mbox{Inv-Gamma}\, ((n-1)/2,\hspace{0.1em} (n-1)s^2/2)
\end{equation}

All the main results that have just been outlined were previously given with more explanation in
Bowater~(2019a), and indeed, a similar derivation of these results can be found in Bowater~(2018a).
By contrast, in what follows, the results that will be presented are generally original results,
i.e.\ results not discussed in earlier papers, although various references will be made to examples
that have been detailed previously.

In the scenario currently being considered, let us now turn our attention to the case where we have
important pre-data knowledge about either of the parameters $\mu$ or $\sigma^2$ that can be
adequately represented by a probability density function over the parameter concerned conditional
on the other parameter being known.
To give an example, let us assume that our pre-data opinion about $\sigma^2$ conditional on $\mu$
being known can be ad-{\linebreak}equately represented by the density function of $\sigma^2$
conditional on $\mu$ that is defined by:
\begin{equation}
\label{equ24}
\sigma^2\,|\,\mu \sim \mbox{Inv-Gamma}\, (\alpha_0, \beta_0)
\end{equation}
where $\alpha_0>0$ and $\beta_0>0$ are given constants.
Treating this density function as a prior density function, and combining it with the likelihood
function in this case, under the Bayesian paradigm, leads to a posterior density of $\sigma^2$
conditional on $\mu$ that is \linebreak defined by:
\pagebreak
\begin{equation}
\label{equ20}
\sigma^2\,|\,\mu,x \sim \mbox{Inv-Gamma}\, (\alpha_0 + (n/2),\hspace{0.1em} \beta_0 +
(n\bm\hat{\sigma}^2/2))
\vspace{1ex}
\end{equation}

If there was no or very little pre-data knowledge about $\mu$, then it would be quite natural to
let the full conditional fiducial density $f(\mu\,|\,\sigma^2,x)$ defined by
equation~(\ref{equ11}), and the full conditional posterior density $p(\sigma^2\,|\,\mu,x)$ defined
in the equation just given, form the basis for using the framework described in Section~\ref{sec2}
to determine the joint post-data density of $\mu$ and $\sigma^2$, i.e.\ the density
$p(\mu, \sigma^2\,|\,x)$.
In fact, by using the simple analytical method outlined in the opening part of Section~\ref{sec2},
it can be easily established that these full conditional densities are compatible, and it is clear
that the joint density function for $\mu$ and $\sigma^2$ that they define must be unique.
This joint density function is therefore the post-data density $p(\mu, \sigma^2\,|\,x)$.
Furthermore, the marginal density of $\mu$ over this joint post-data density is given by:
\vspace{2ex}
\begin{equation}
\label{equ13}
\mu\,|\,x \sim \mbox{Non-standardised}\ t_{2\alpha_0 + n - 1}\hspace{-0.1em}
\left( \bar{x}, \left(\frac{2\beta_0 + (n-1)s^2}{(2\alpha_0 + n - 1)n} \right)^{\!\!0.5\,}
\right),
\vspace{2ex}
\end{equation}
while the marginal density of $\sigma^2$ over the joint density in question is given by:
\begin{equation}
\label{equ14}
\sigma^2\,|\,x \sim \mbox{Inv-Gamma}\, (\hspace{0.05em}\alpha_0 + ((n-1)/2),\hspace{0.15em}
\beta_0 + ((n-1)/2)s^2\hspace{0.05em})
\vspace{0.5ex}
\end{equation}

To illustrate this example, Figure~1 shows some results from using the calculations just described
to perform an analysis of a data set $x$ that is summarised by the values $n=9$, $\bar{x}=2.7$ and
$s^2=9$.
In particular, this figure shows a plot of the specific form of the conditional prior density
$p(\sigma\,|\,\mu)$ as defined by equation~(\ref{equ24}) that was used in this analysis, which is
represented by the short-dashed curve in Figure~1(b), a plot of the marginal post-data density
$p(\mu\,|\,x)$ as defined by equation~(\ref{equ13}), which is represented by the long-dashed
(rather than the dot-dashed) curve in Figure~1(a), and a plot of the marginal post-data density
$p(\sigma\,|\,x)$ as given by equation~(\ref{equ14}), \pagebreak which is represented by the
long-dashed curve in Figure~1(b).
To complete the specification of the prior density $p(\sigma\,|\,\mu)$, the constants $\alpha_0$
and $\beta_0$ in equation~(\ref{equ24}) were set equal to 4 and 64 respectively.
These settings imply that this prior density would be equal to the marginal fiducial density of
$\sigma$ defined by equation~(\ref{equ15}) if this latter density was based on having observed a
variance of 16 in a preliminary sample of 9 observations drawn from a population \linebreak having
the same unknown variance $\sigma^2$ that is currently being considered.
Notice that, from a practical viewpoint, this interpretation would be genuinely useful if the mean
$\mu$ of this population was not only assumed to be unknown, but was assumed not to be the same as
the mean $\mu$ of present interest.

On the basis of only the main data set being analysed, i.e.\ the data set $x$, and for comparison
with the plots being considered, the solid curves in Figures~1(a) and~1(b) represent, respectively,
the marginal fiducial density $f(\mu\,|\,x)$ as defined by equation~(\ref{equ16}) and the marginal
fiducial density $f(\sigma\,|\,x)$ as given by equation~(\ref{equ15}).

Let us now change the state of knowledge about both the parameters $\mu$ and $\sigma^2$ before the
data were observed. In particular, let us begin by imagining that we have important pre-data
knowledge about the mean $\mu$ that can be adequately represented by a probability density function
over $\mu$ conditional on $\sigma^2$ being known, i.e.\ the density $p(\mu\,|\,\sigma^2)$.
\linebreak To give an example, let this density function be defined by:
\vspace{0.25ex}
\begin{equation}
\label{equ25}
\mu\,|\,\sigma^2 \sim \mbox{Non-standardised}\ t_{\nu_0}(\mu_0, \sigma_0)
\vspace{0.25ex}
\end{equation}
where $\nu_0 > 0$, $\sigma_0 > 0$ and $\mu_0$ are given constants.
Treating this choice of the density $p(\mu\,|\,\sigma^2)$ as a prior density under the Bayesian
paradigm leads to a posterior density of $\mu$ conditional on $\sigma^2$ that is defined by:
\vspace{0.5ex}
\[
p(\mu\,|\,\sigma^2,x) \propto (1 + (1/\sigma_0^2\nu_0)(\mu - \mu_0)^2)^{-(\nu_0+1)/2}
\exp (-n(\bar{x} - \mu)^2/2\sigma^2)
\vspace{0.5ex}
\]

If now we assume that there was no or very little pre-data \pagebreak knowledge about $\sigma^2$,
then it would be quite natural to use the full conditional fiducial density $f(\sigma^2\,|\,\mu,x)$
given by equation~(\ref{equ17}), and the full conditional posterior density $p(\mu\,|\,\sigma^2,x)$
defined by the equation just presented, as the basis for determining the joint post-data density of
$\mu$ and $\sigma^2$, i.e.\ the density $p(\mu, \sigma^2\,|\,x)$.
Similar to the previous example, it can easily be shown by using once again the simple analytical
method outlined in the opening part of Section~\ref{sec2} that these full conditional densities are
compatible, and it is again clear that the joint density function for $\mu$ and $\sigma^2$ that
they define must be unique.
This joint den\-sity function, which is therefore the post-data density $p(\mu, \sigma^2\,|\,x)$,
can in fact be ex-{\linebreak}pressed as follows:
\vspace{0.5ex}
\begin{equation}
\label{equ18}
p(\mu,\sigma^2\,|\,x) = (1/\sigma^2)^{(n/2)+1}(1 + (1/\sigma_0^2\nu_0)(\mu -
\mu_0)^2)^{-(\nu_0+1)/2} \exp (-(1/2\sigma^2)n\bm\hat{\sigma}^2)
\vspace{1ex}
\end{equation}

\begin{figure}[t]
\begin{center}
\noindent
\makebox[\textwidth]{\includegraphics[width=7.25in]{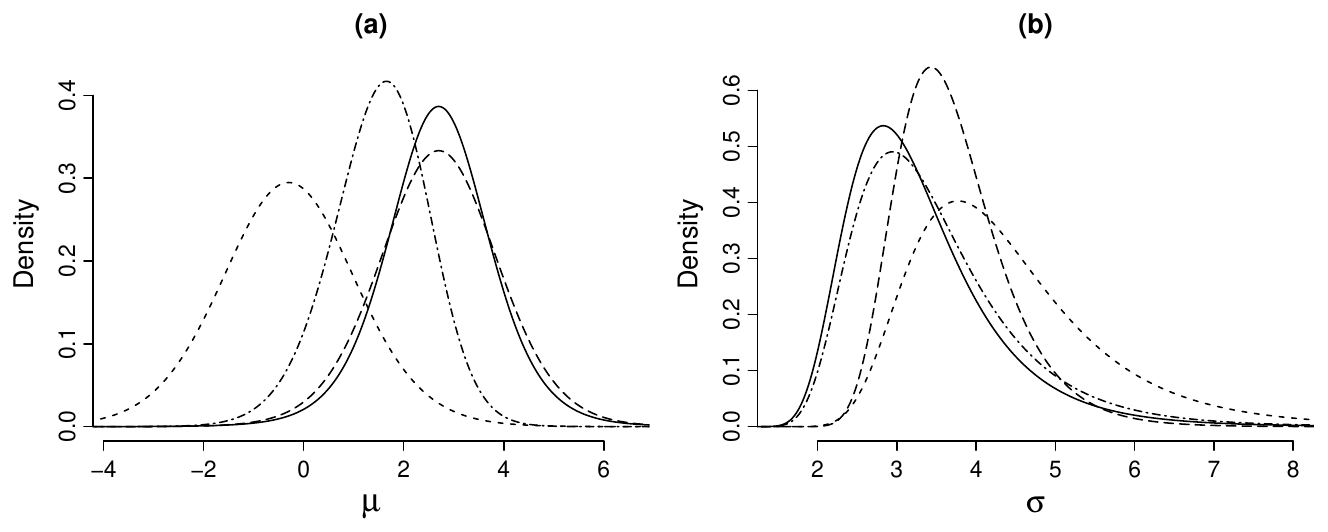}}
\caption{\small{Conditional prior and marginal post-data densities of the mean $\mu$ and standard
deviation $\sigma$ of a normal distribution}}
\end{center}
\end{figure}

To illustrate the use of the method being discussed, let us apply this method to the analysis of
the same data set $x$ as we were concerned with in the previous example. In particular, Figure~1
shows, along with the plots that were \pagebreak mentioned earlier, a plot of the specific form of
the conditional prior density $p(\mu\,|\,\sigma^2)$ as defined by equation~(\ref{equ25}) that was
used in the present analysis, which is represented by the short-dashed curve in Figure~1(a), and
plots of the marginal densities of $\mu$ and $\sigma$ over the joint post-data density
$p(\mu, \sigma^2\,|\,x)$ given in equation~(\ref{equ18}), which are represented by the dot-dash
curves in Figure~1(a) and Figure~1(b) respectively.
These marginal densities of $\mu$ and $\sigma$ were obtained by numerical integration over the
joint density $p(\mu, \sigma^2\,|\,x)$.
To complete the specification of the prior density $p(\mu\,|\,\sigma^2)$, the constants in
equation~(\ref{equ25}) were given the settings $\nu_0=17$, $\mu_0=-0.3$ and $\sigma_0=4/3$.
These settings imply that this prior density would be equal to the marginal fiducial density of
$\mu$ given by equation~(\ref{equ16}) if this latter density was based on having observed a mean of
$-0.3$ and a variance of 32 in a prelim-{\linebreak}inary sample of 18 observations drawn from a
population having the same unknown mean $\mu$ that is currently being considered.
Similar to a point made earlier, such an interpretation would be genuinely useful in a practical
sense if the variance $\sigma^2$ of this population was not only assumed to be unknown, but was
assumed not to be the same as the variance $\sigma^2$ of present interest.

Finally, in the case where we have important pre-data knowledge about both $\mu$ and $\sigma^2$
that can be adequately represented by full conditional probability densities over each of these
parameters, i.e.\ the densities $p(\mu\,|\,\sigma^2)$ and $p(\sigma^2\,|\,\mu)$, it would seem
reasonable, assuming that these conditional densities are compatible, to treat these densities as
being conditional prior densities, and to use exclusively the standard Bayesian approach to make
inferences about $\mu$ and $\sigma^2$. Since Bayesian inference is a well-known form of inference,
no further discussion of this particular case will be given here.

\vspace{3ex}
\subsection{Alternative solution to Student's problem}
\label{sec12}

In the previous section, Student's problem was tackled by incorporating organic fiducial inference
and Bayesian inference into the framework outlined in Section~\ref{sec2}, now let us consider a
case in which it would seem appropriate to address the same problem by also incorporating bispatial
inference into this framework.

In particular, let us assume that conditional on the variance $\sigma^2$ being known, the scenario
of interest of Section~\ref{sec5} would apply if the general parameter $\theta_j$ was taken as
being the mean $\mu$, with the interval $[\theta_{j0}, \theta_{j1}]$ in this scenario being denoted
now as the interval $[\hspace{0.05em}\mu_{1}-\varepsilon,
\mu_{1}+\varepsilon\hspace{0.05em}]$, where $\varepsilon \geq 0$ and $\mu_{1}$ are given constants.
We will therefore construct the post-data density of $\mu$ conditional on $\sigma^2$ using the type
of bispatial infer-{\linebreak}ence described in Section~\ref{sec5}.

To do this, the test statistic $T(x)$ as defined in Section~\ref{sec5} will be quite reasonably
assumed to be the sample mean $\bar{x}$. Therefore, in the case where the mean $\bar{x}$ is greater
than zero, which will be assumed to be the case of particular interest, the hypotheses $H_{P}$ and
$H_{S}$ will be as defined in equations~(\ref{equ10}) and~(\ref{equ7}), which implies that, for the
present example, they can be more specifically expressed as:
\vspace{0.5ex}
\begin{gather}
H_{P}: \mu \leq \mu_{1} + \varepsilon \nonumber\\
H_{S}: \rho (\overbar{X}^{*} > \bar{x}) \leq 1 - \Phi((\bar{x} - \mu_{1} - \varepsilon) \sqrt{n}/
\sigma)\ \ \ (=\pvalue) \label{equ21}
\end{gather}

\vspace{0.5ex}
\noindent
where $\overbar{X}^{*}$ is the mean of an as-yet-unobserved sample of $n$ additional values drawn
from the density function $g(x\,|\,\mu,\sigma^2)$, i.e.\ the normal density function being studied,
and $\Phi(y)$ is the cumulative density of a standard normal distribution at the value $y$.
Also, it will be assumed, quite reasonably, that the fiducial density $f_{S}(\theta_j\,|\, x)$,
which is required by equations~(\ref{equ8}) and~(\ref{equ19}), i.e.\ the density
$f_{S}(\mu\,|\,\sigma^2,x)$ in the present case, is the fiducial density of $\mu$ given $\sigma^2$
that was defined in equation~(\ref{equ11}).

\begin{figure}[t]
\begin{center}
\noindent
\makebox[\textwidth]{\includegraphics[width=7.25in]{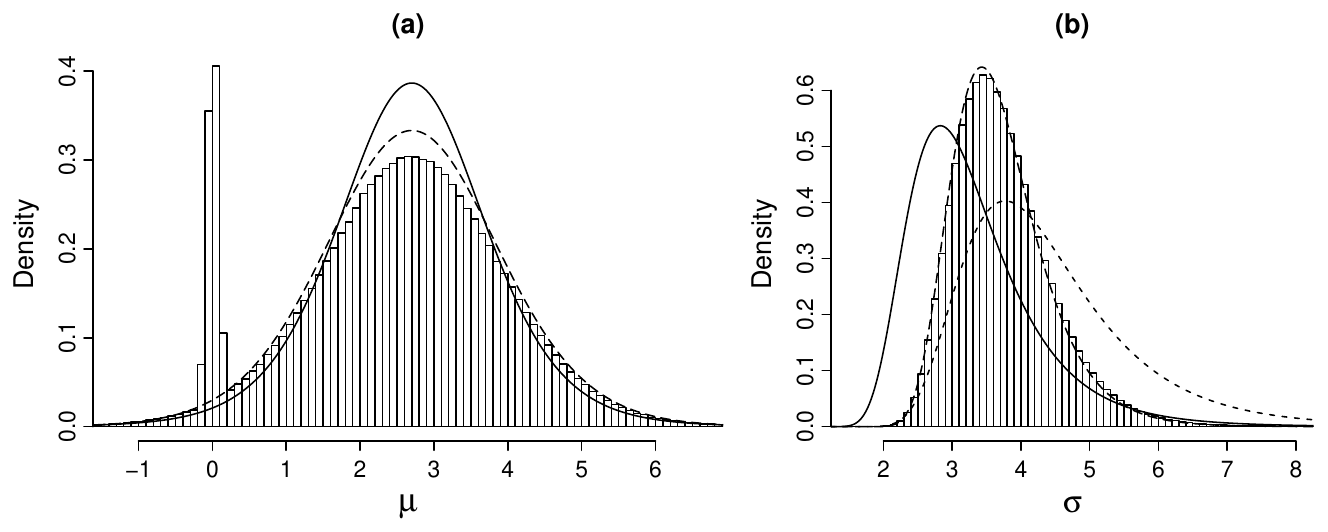}}
\caption{\small{Histograms representing marginal post-data densities of the mean $\mu$ and
standard deviation $\sigma$ of a normal distribution}}
\end{center}
\end{figure}

To complete the specification of the post-data density of $\mu$ given $\sigma^2$, i.e.\ in keeping
with earlier notation, the density $b(\mu\,|\,\sigma^2,x)$, let us now make some more specific
assumptions.
In particular, let us assume that $\mu_1=0$ and $\varepsilon=0.2$, and that the den\-sity function
$h(\theta_j)$ that appears in equation~(\ref{equ9}), i.e.\ the density \pagebreak $h(\mu)$ in the
present \linebreak case, is defined by:
\vspace{0.75ex}
\begin{equation}
\label{equ35}
\mu \sim \mbox{Beta}\hspace{0.1em} (4,4,-0.2,0.2)
\vspace{0.75ex}
\end{equation}
i.e.\ it is a beta density function for $\mu$ on the interval $[-0.2,0.2]$ with both its shape
parameters equal to 4.
Furthermore, we will assume that the data is summarised as it was in the previous section, i.e.\ by
$n=9$, $\bar{x}=2.7$ and $s^2=9$.
Finally, the probabilities $\kappa$ that would be assigned to the hypothesis $H_{S}$ in
equation~(\ref{equ21}) for different values of $\sigma^2$ will be assumed to be given by the PDO
curve for $\mu$ conditional on $\sigma^2$ that has the formula:
$\kappa = \pvalue^{\hspace{0.1em}0.6}$, where, as indicated in equation~(\ref{equ21}), $\pvalue$ is
the one-sided P value in the definition of the hypothesis $H_{S}$ concerned.
These assumptions fully specify the post-data density $b(\mu\,|\,\sigma^2,x)$ according to the
methodology outlined in Section~\ref{sec5}.

In fact, in Bowater~(2019b), this full conditional density of $\mu$, precisely as this density has
just been defined, and the full conditional fiducial density $f(\sigma^2\,|\,\mu, x)$ given by
equation~(\ref{equ17}), with the data set $x$ assumed to be as currently specified, \pagebreak were
used as the basis for determining the joint post-data density of $\mu$ and $\sigma^2$ within the
same type of frame\-work as described in Section~\ref{sec2}.
As mentioned earlier, the use of the full conditional fiducial density of $\sigma^2$ being referred
to would be quite natural if it was assumed there was no or very little pre-data knowledge about
the variance $\sigma^2$. However, this assumption will not be made here.
Instead, let us assume that we have important pre-data knowledge about $\sigma^2$ that in fact is
adequately represented by the density function for $\sigma^2$ conditional on $\mu$ that is defined
by equation~(\ref{equ24}), with the same choices for the con\-stants $\alpha_0$ and $\beta_0$ as
were used earlier to express pre-data knowledge about $\sigma^2$ conditional on $\mu$, i.e.\ with
$\alpha_0=4$ and $\beta_0=64$.
Treating this density function as a prior density function under the Bayesian paradigm leads
therefore to the posterior density of \linebreak $\sigma^2$ given $\mu$,  i.e.\ the density
$p(\sigma^2\,|\,\mu,x)$, being defined as it was in equation~(\ref{equ20}).

To illustrate this example, Figure~2 shows some results from running a Gibbs sampler on the basis
of the full conditional post-data densities of $\mu$ and $\sigma^2$ that have just been defined,
i.e.\ the post-data density $b(\mu\,|\,\sigma^2,x)$ and the posterior density
$p(\sigma^2\,|\,\mu,x)$, with a uniform random scanning order of the parameters $\mu$ and
$\sigma^2$, as such a scanning order \linebreak was defined in Section~\ref{sec2}.
In particular, the histograms in Figures~2(a) and~2(b) represent the distributions of the values of
the mean $\mu$ and the standard deviation $\sigma$, respectively, over a single run of six million
samples of these parameters generated by the Gibbs sam\-pler after a preceding run of two thousand
samples, which were classified as belonging to its burn-in phase, had been discarded.
The sampling of the density $b(\mu\,|\,\sigma^2,x)$ was based on the Metropolis algorithm
(Metropolis et al.~1953), while each value drawn from the density $p(\sigma^2\,|\,\mu,x)$ was
independent from the preceding iterations.

In addition to this analysis, the Gibbs sampler was also run various times from different starting
points, and a careful study of the output of these runs using appropriate diagnostics provided no
evidence to suggest that the sampler does not have a limiting distribution, and showed, at the
same time, that it would appear to generally converge quickly to this distribution.
Furthermore, the Gibbs sampling algorithm was run separately with each of the two possible fixed
scanning orders of the parameters, i.e.\ the one in which $\mu$ is updated first and then
$\sigma^2$ is updated, and the one that has the reverse order, in accordance with how a single
transition of such an algorithm was defined in Section~\ref{sec2}, i.e.\ single transitions of the
algorithm incorporated updates of both parameters.
In doing this, no statistically significant difference was found between the samples of parameter
values aggregated over the runs of the sampler in using each of these two scanning orders after
excluding the burn-in phase of the sampler, e.g.\ between the two sample correlations of $\mu$ and
$\sigma$, even when the runs concerned were long.
Taking into account what was discussed in Section~\ref{sec2}, this implies that the full
conditional densities of the limiting distribution of the original Gibbs sampler, i.e.\ the one
with a uniform random scanning order, should be, at the very least, close approximations to the
full conditional densities on which the sampler is based, i.e.\ the post-data density
$b(\mu\,|\,\sigma^2,x)$ and the posterior density $p(\sigma^2\,|\,\mu,x)$ defined earlier.

Each of the curves overlaid on the histograms in Figures~2(a) and~2(b), which are distinguished by
being plotted with short-dashed, long-dashed and solid lines, is identical to the curve plotted
using the same line type in Figures~1(a) and~1(b) respectively.
By comparing the histograms in Figures~2(a) and~2(b) with the curves in question, it can be seen
that the forms of the marginal post-data densities of $\mu$ and $\sigma$ that are represented by
these histograms are consistent with what we would have intuitively expected given the pre-data
beliefs about $\mu$ and $\sigma$ that have been taken into account.
It may also be to some extent informative to compare Figures~2(a) and~2(b) with Figures~4(a)
and~4(b) of Bowater~(2019b), since these latter figures relate to the example from this earlier
paper that was mentioned midway through the present section.

\vspace{3ex}
\subsection{Inference about a trinomial distribution}
\label{sec11}

We will now consider the problem of making inferences about the parameters
$\pi =(\pi_1,\linebreak \pi_2, \pi_3)'$ of a trinomial distribution, where $\pi_{i}$ is the
proportion of times that the $i$th outcome of the three possible outcomes is generated in the long
run, based on observing a sample of counts $x=(x_1, x_2, x_3)'$ from the distribution concerned,
where $x_i$ is the num\-ber of times that the $i$th outcome is observed. Since of course
$\pi_1+\pi_2+\pi_3=1$, this model has effectively only two parameters, which we will assume to be
the proportions $\pi_1$ and $\pi_2$. To clarify, the probability of observing the sample of counts
$x=(x_1, x_2, x_3)'$ is specified by the trinomial mass function in this case, i.e.\ the function:
\vspace{2ex}
\[
g_0(x\,|\,\pi_1,\pi_2) =
\left\{
\begin{array}{ll}
(n!/x_{1}!\hspace{0.05em}x_{2}!\hspace{0.05em}x_{3}!) \pi_{1}^{x_1} \pi_2^{x_2} \pi_3^{x_3}\ &
\mbox{if $x_1,x_2,x_3 \in \mathbb{Z}_{\geq 0}$ and
$n = x_1\hspace{-0.05em}+\hspace{-0.05em}x_2\hspace{-0.05em}+\hspace{-0.05em}x_3$}\\[1ex]
0 & \mbox{otherwise}
\end{array}
\right.
\vspace{2ex}
\]
where the total number of counts $n$ is fixed.

In particular, let us begin by applying organic fiducial inference as outlined in
Section~\ref{sec4} to make inferences about $\pi_2$ conditional on $\pi_1$ being known. In this
regard, observe that if $\pi_1$ was known, sufficient statistics for $\pi_2$ would be $x_2$ and
$x_2+x_3$. However, $x_2+x_3$ is an ancillary complement of $x_2$, and therefore, according to the
more general definition of the fiducial statistic $Q(x)$ given in Bowater~(2019a), the count $x_2$
can justifiably be assumed to be the statistic $Q(x)$.
Based on this assumption and given a value for $\pi_1$, equation~(\ref{equ3}) can naturally be
redefined as:
\vspace{1ex}
\begin{equation}
\label{equ46}
x_2=\varphi(\Gamma,\pi_2)= \min \left\{ y: \Gamma < \mbox{\large $\sum$}_{\mbox{\footnotesize
$j\hspace{-0.25em}=\hspace{-0.25em}0$}}^{\mbox{\footnotesize $y$}}\hspace{0.3em}
g_1(j\,|\,\pi_2) \right\}
\vspace{1ex}
\end{equation}
where the primary r.v.\ $\Gamma$ has a uniform distribution over the interval $(0,1)$, and the
function $g_1(j\,|\,\pi_2)$ is given by:
\pagebreak
\[
g_1(j\,|\,\pi_2) = \frac{(x_2+x_3)!}{(x_2+x_3-j)!\hspace{0.05em}j!}  \left( \frac{\pi_2}{1 -
\pi_1} \right)^{\hspace{-0.05em}j} \left( \frac{1-\pi_1-\pi_2}{1 - \pi_1}
\right)^{\hspace{-0.05em}x_2 + x_3 - j}
\vspace{2ex}
\]
in which the statistic $x_2+x_3$ is treated as having already been generated.

Given that it will be assumed that there was no or very little pre-data knowledge about the
proportion $\pi_2$, the GPD function for $\pi_2$ will be quite reasonably specified as follows:
$\omega_G(\pi_2) = a$ if $0 \leq \pi_2 \leq 1-\pi_1$ and 0 otherwise, where $a>0$.
However, since for whatever choice is made for this GPD function and whatever turns out to be the
sample $x$, equation~(\ref{equ46}) will never satisfy Condition~1 of Section~\ref{sec4}, the
principle outlined in this earlier section for deriving the fiducial density
$f(\theta_j\,|\,\theta_{-j},x)$ can not be employed in the case of interest to determine the
fiducial density of $\pi_2$ given $\pi_1$, i.e.\ the density $f(\pi_2\,|\,\pi_1,x)$.
This density can instead, though, be determined by applying Principle~2 of Bowater~(2019a), which
as mentioned in Section~\ref{sec4}, is a principle that relies on the concept of a local pre-data
(LPD) function.
In particular, to make use of this principle in the present case, we need to specify a LPD
function for $\pi_2$. Further details about how the principle in question is applied are given in
Bowater~(2019a).

As also discussed in this earlier paper, the type of method being considered could be used to
obtain a complete set of full conditional fiducial densities for $k$ of the population proportions
of a multinomial distribution with $k+1$ categories on the basis of a given sample from this
distribution, which could then be used to determine a joint fiducial density of these $k$
proportions (or equivalently of all $k+1$ population proportions of the distribution) using the
type of framework outlined in Section~\ref{sec2} of the current paper. In relation to this issue, a
detailed example was presented in Bowater~(2019a) of how a joint fiducial density of the five (or
equivalently four of the five) population proportions of a multinomial distribution with five
categories could be obtained using such an approach.

However, in the present case, it will be assumed that, unlike the post-data density of $\pi_2$
given $\pi_1$, the post-data density of $\pi_1$ given $\pi_2$ does not belong \pagebreak to the
class of full conditional fiducial densities under discussion.
This is because, in contrast to the kind of scenario where the type of approach just mentioned is
most applicable, it will be assumed that we have important pre-data knowledge about the proportion
$\pi_1$, and that this pre-data knowledge can, in particular, be adequately represented by a
probability density function over $\pi_1$ conditional on $\pi_2$ being known, i.e.\ the density
$p(\pi_1\,|\,\pi_2)$. To give an example, let this density function be defined by:
\vspace{2ex}
\begin{equation}
\label{equ22}
p(\pi_1\,|\,\pi_2) =
\left\{
\begin{array}{ll}
\mathtt{C}_4 (\pi_1)^{\alpha-1} (1 - \pi_1)^{\beta-1} \ &
\mbox{if $0 \leq \pi_1 \leq 1 - \pi_2$}\\[1ex]
0 & \mbox{otherwise}
\end{array}
\right.
\vspace{2ex}
\end{equation}
where $\alpha>0$ and $\beta>0$ are given constants, and $\mathtt{C}_4$ is a normalising constant.
Treating this choice of the density $p(\pi_1\,|\,\pi_2)$ as a prior density and combining it with
the likelihood function in this case, under the Bayesian paradigm, leads to a posterior density of
\linebreak $\pi_1$ given $\pi_2$ that is defined by:
\vspace{2ex}
\[
p(\pi_1\,|\,\pi_2,x) =
\left\{
\begin{array}{ll}
\mathtt{C}_5 (\pi_1)^{\alpha+x_1-1} (1-\pi_1-\pi_2)^{n-x_1-x_2} (1 - \pi_1)^{\beta-1}\ &
\mbox{if $0 \leq \pi_1 \leq 1 - \pi_2$}\\[1ex]
0 & \mbox{otherwise}
\end{array}
\right.
\vspace{2ex}
\]
where $\mathtt{C}_5$ is a normalising constant.

To illustrate this example, Figure~3 shows some results from running a Gibbs sampler on the basis
of the full conditional post-data densities of $\pi_1$ and $\pi_2$ that have just been referred to,
i.e.\ the fiducial density $f(\pi_2\,|\,\pi_1,x)$ and the posterior density (derived using Bayesian
inference) $p(\pi_1\,|\,\pi_2,x)$, with a uniform random scanning order of the parameters $\pi_1$
and $\pi_2$.
In particular, the histograms in Figures~3(a) and~3(b) represent the distributions of the values of
$\pi_1$ and $\pi_2$, respectively, over a single run of six million samples of these parameters
generated by the Gibbs sampler after a preceding run of one thousand samples were discarded due to
these samples being classified as belonging to its burn-in phase.
The sampling of the density $p(\pi_1\,|\,\pi_2,x)$ \pagebreak was based on the Metropolis
algorithm, while the sampling of the density $f(\pi_2\,|\,\pi_1,x)$ was independent from the
preceding iterations.

\begin{figure}[t]
\begin{center}
\noindent
\makebox[\textwidth]{\includegraphics[width=7.25in]{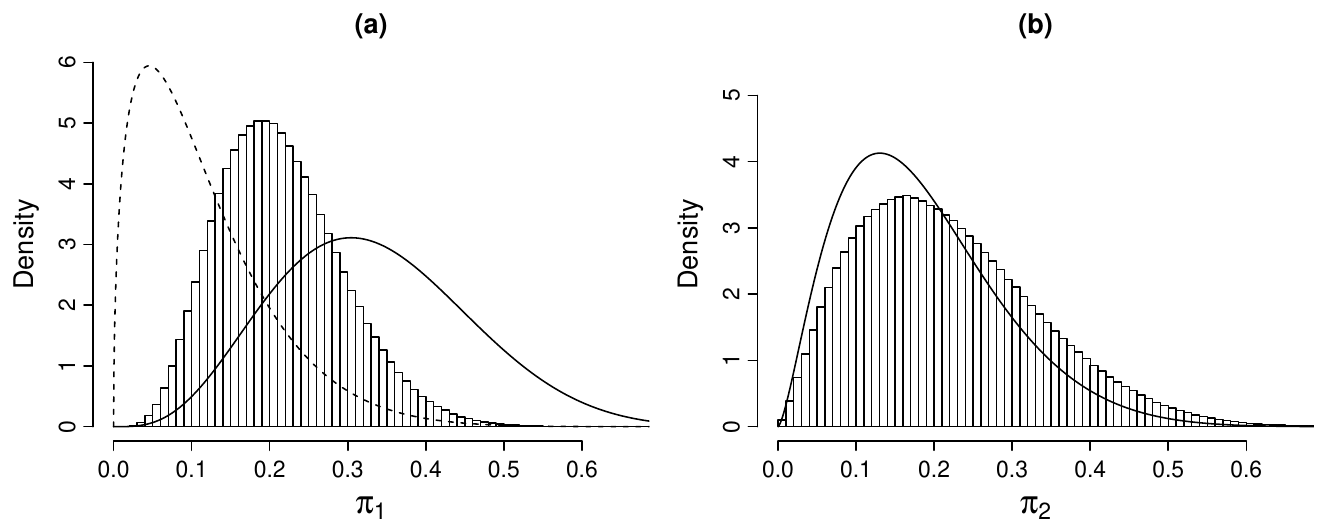}}
\caption{\small{Unconditional prior density of one parameter, namely $\pi_1$, and marginal
post-data densities of both parameters $\pi_1$ and $\pi_2$ of a trinomial distribution}}
\end{center}
\end{figure}

Moreover, the observed counts on which the inferential process being described was based were set
as follows: $x_1=4$, $x_2=2$ and $x_3=6$. Also, it was assumed that the LPD function for $\pi_2$
was given by:
\vspace{1.5ex}
\[
\omega_L(\pi_2) = \left\{
\begin{array}{ll}
b \ & \mbox{if $0 \leq \pi_2 \leq 1-\pi_1$}\\[1ex]
0 & \mbox{otherwise}
\end{array}
\right.
\vspace{2ex}
\]
where $b>0$, which is in keeping with the choices that were made for functions of this kind in the
aforementioned example in Bowater~(2019a) of the use of organic fiducial inference in this type of
situation.
Finally, the specification of the prior density\hspace{0.03em} $p(\pi_1\hspace{0.12em}|\,\pi_2)$
was completed by making the assignments $\alpha=1.5$ and $\beta=11.5$ in equation~(\ref{equ22}).

Observe that these choices for the variables $\alpha$ and $\beta$ imply that the prior density
$p(\pi_1\,|\,\pi_2)$ is equal to the density function of $\pi_1$ \pagebreak that is defined by:
\begin{equation}
\label{equ23}
p(\pi_1) \propto (\pi_1)^{0.5} (1 - \pi_1)^{10.5} \ \ \ \mbox{if $0 \leq \pi_1 \leq 1$ and
equal to $0$ otherwise}
\vspace{0.5ex}
\end{equation}
conditioned on the inequality $\pi_1 \leq 1-\pi_2$, which clearly must always hold, but is of
course a condition that can only be applied if the proportion $\pi_2$ is known.
Furthermore, this latter unconditioned density $p(\pi_1)$ is equivalent to the (unconditional)
posterior density of $\pi_1$ that would be formed after observing the counts $x_1=1$ and
$x_2+x_3=11$ (for which, we can see, membership of categories 2 and 3 is not distinguished) if the
prior density of $\pi_1$ was the Jeffreys prior that corresponds to conducting the binomial
experiment that produced these counts (see Jeffreys~1961).
However, since as mentioned in Section~\ref{sec3}, posterior densities formed on the basis of prior
densities that are depend\-ent on the sampling model, such as the Jeffreys prior, are
controversial, it is arguably of more interest to note that this posterior density of $\pi_1$ is a
close approximation to forms of the (unconditional) fiducial density of $\pi_1$ that would be
naturally constructed on the basis of the two counts in question, i.e.\ $x_1=1$ and $x_2+x_3=11$,
by applying the methodology in Bowater~(2019a) if nothing or very little was known about the
proportion $\pi_1$ before these counts were observed.
This type of approximation was discussed both in \linebreak this previous paper and in
Bowater~(2019b).

In addition to the analysis just described, the Gibbs sampler of present interest was also run
various times from different starting points, and there was no suggestion from using appropriate
diagnostics that the sampler does not have a limiting distribution.
Furthermore, after excluding the burn-in phase of the sampler, no statistically significant
difference was found between the samples of parameter values aggregated over the runs of the
sampler in using each of the two fixed scanning orders of the parameters $\pi_1$ and $\pi_2$ that
are possible, with a single transition of the sampler defined in the same way as in the example
outlined in the previous section, even when the runs concerned were long.
Therefore, taking into account what was discussed in Section~\ref{sec2}, the full conditional
densities of the limiting distribution of the original random-scan Gibbs sampler should be, at the
very least, close approximations to the full conditional densities on which the sampler is based,
i.e.\ the posterior density $p(\pi_1\,|\,\pi_2,x)$ and the fiducial den-{\linebreak}sity
$f(\pi_2\,|\,\pi_1,x)$ defined earlier.

The solid curves overlaid on the histograms in Figures~3(a) and~3(b) are plots of the marginal
densities of the parameters $\pi_1$ and $\pi_2$, respectively, over the joint posterior density of
$\pi_1$ and $\pi_2$ that would be formed after having only observed the main data of interest,
i.e.\ the counts $x_1=4$, $x_2=2$ and $x_3=6$, if the joint prior density of these parameters was
the Jeffreys prior for this case. It can be shown that this joint posterior density, which is in
fact defined by the expression:
\vspace{2ex}
\[
p(\pi_1, \pi_2\,|\,x)\hspace{-0.05em} = \hspace{-0.05em}\left\{\hspace{-0.1em}
\begin{array}{ll}
\mathtt{C}_6 (\pi_1)^{x_1-0.5} (\pi_2)^{x_2-0.5} (1-\pi_1-\pi_2)^{x_3-0.5} \
& \mbox{if $\pi_1, \pi_2 \in [0,1]$ and $\pi_1+\pi_2 \leq 1$}\\[1ex]
0 & \mbox{otherwise}
\end{array}
\right.
\vspace{2ex}
\]
where $\mathtt{C}_6$ is a normalising constant, is a close approximation to forms of the joint
fiducial density of $\pi_1$ and $\pi_2$ that would be naturally constructed on the basis of these
observed counts $x_1$, $x_2$ and $x_3$ by applying the methodology in Bowater~(2019a) if there was
no or very little pre-data knowledge about $\pi_1$ and $\pi_2$.
The dashed curve overlaid on the histogram in Figure~3(a) is a plot of the density function of
$\pi_1$ given in equation~(\ref{equ23}), i.e.\ the unconditioned prior density $p(\pi_1)$.

By comparing the locations and degrees of dispersion of the histograms in Figures~3(a) and~3(b), it
can be seen that it is beyond dispute that generally more precise conclusions can be drawn about
the proportion $\pi_1$ than the proportion $\pi_2$ after the counts $x_1$, $x_2$ and $x_3$ in
question have been observed, which, on the basis of comparing these histograms with the curves
overlaid on them, can be clearly attributed to the incorporation, under the Bayesian paradigm, of
substantial prior information about $\pi_1$ into the construction of the joint post-data density of
$\pi_1$ and $\pi_2$.

\pagebreak
\subsection{Inference about a linear regression model}
\label{sec13}

Let us now turn our attention to the problem of making inferences about all the parameters
$\beta_0$, $\beta_1$, $\beta_2$, $\beta_3$ and $\sigma^2$ of the normal linear regression model
defined by:
\vspace{0.5ex}
\begin{equation}
\label{equ26}
Y = \beta_0 + \beta_1 x_{1} + \beta_2 x_{2} + \beta_3 x_{3} + \varepsilon
\ \ \ \, \mbox{with $\varepsilon \sim \mbox{N}(0,\sigma^2)$}
\vspace{0.5ex}
\end{equation}
where $Y$ is the response variable and $x_{1}$, $x_{2}$ and $x_{3}$ are three covariates, on the
basis of a data set $y_{+}=\{(y_i, x_{1i}, x_{2i}, x_{3i}):i=1,2,\ldots,n\}$, where $y_i$ is the
value of\hspace{0.03em} $Y$\hspace{-0.03em} generated by this model for the $i$th case in this data
set given values $x_{1i}$, $x_{2i}$ and $x_{3i}$ of the covariates $x_{1}$, $x_{2}$ and $x_{3}$
respectively.

Observe that sufficient statistics for each of the parameters $\beta_0$, $\beta_1$, $\beta_2$,
$\beta_3$ and $\sigma^2$ conditional on all parameters except the parameter itself being known are
respectively:
\vspace{1.75ex}
\begin{equation}
\label{equ27}
\hspace{-0.3em}\sum_{i=1}^{n} y_i,\ \sum_{i=1}^{n} x_{1i} y_i,\ \sum_{i=1}^{n} x_{2i} y_i,
\ \sum_{i=1}^{n} x_{3i} y_i\ \ \mbox{and}\ \,
\sum_{i=1}^{n} (y_i - \beta_0 - \beta_1 x_{1i} - \beta_2 x_{2i} - \beta_3 x_{3i})^2
\vspace{2.25ex}
\end{equation}
In Bowater~(2018a), all except the fourth statistic here were used as fiducial statistics
$Q(y_{+})$ to derive, under the strong fiducial argument, a complete set of full conditional
fiducial densities of the model parameters in the special case where the model in
equation~(\ref{equ26}) is a quadratic regression model, i.e.\ where $x_{2}=(x_{1})^2$ and the
coefficient $\beta_3$ is set to zero (hence the lack of a need for the fourth statistic).
Also, it was shown in this earlier paper that, since these full conditional densities are
compatible, they directly define a unique joint density for $\beta_0$, $\beta_1$, $\beta_2$ and
$\sigma^2$, which is therefore a joint fiducial density for these parameters.
Furthermore, it is fairly clear from this previous analysis how the particular method of inference
that was employed can be extended to address the problem of making inferences about the parameters
of the more general type of normal linear regression model that is \pagebreak defined by
equation~(\ref{equ26}).

However, this specific type of method is not going to be directly applicable to the case that will
be presently considered. This is because, although it will be assumed that nothing or very little
was known about the parameters $\beta_0$, $\beta_2$ and $\sigma^2$ before the data were observed,
by contrast it is going to be assumed that there was a substantial amount of pre-data knowledge
about the parameters $\beta_1$ and $\beta_3$. Let us begin though by clarifying how the full
conditional post-data densities of $\beta_0$, $\beta_2$ and $\sigma^2$ will be constructed.

With this aim in mind, notice that if the sufficient statistics for $\beta_0$ and $\beta_2$
presented in equation~(\ref{equ27}) are treated as the fiducial statistics $Q(y_{+})$ in making
inferences about these two parameters respectively, then given that the sampling distributions of
these statistics are normal, the functions $\varphi(\Gamma,\beta_0)$ and $\varphi(\Gamma,\beta_2)$,
as generally defined by equation~(\ref{equ3}), can be expressed in a similar way to how the
function $\varphi(\Gamma,\mu)$ was expressed in equation~(\ref{equ28}).
Also if, under the condition that $\sigma^2$ is the only unknown parameter, the sufficient
statistic for $\sigma^2$ presented in equation~(\ref{equ27}) is treated as the statistic $Q(y_{+})$
in making inferences about this parameter, then given that this statistic divided by $\sigma^2$ has
a chi-squared sampling distribution with $n$ degrees of freedom, the function
$\varphi(\Gamma,\sigma^2)$ can be expressed in a similar way to how this type of function was
expressed in equation~(\ref{equ29}), where it was also denoted as $\varphi(\Gamma,\sigma^2)$ but
with of course a different meaning.
Furthermore, given what has been assumed, it would be quite natural to specify the GPD function for
$\sigma^2$ in the same way as the GPD function for a population variance (also denoted as
$\sigma^2$) was defined in equation~(\ref{equ30}), and to specify the GPD \linebreak functions for
$\beta_0$ and $\beta_2$ as follows: $\omega_G(\beta_i) = a$ for $\beta_i \in (-\infty,\infty)$,
where $a>0$.
This leads to the full conditional fiducial densities for $\beta_0$, $\beta_2$ and $\sigma^2$ being
defined as follows:
\vspace{0.5ex}
\begin{gather}
\beta_0\,|\,\beta_{-0},\sigma^2,y_{+} \sim \mbox{N}\hspace{-0.1em} \left(\hspace{0.1em}
\mbox{$\sum_{i=1}^{n}$}\hspace{0.1em} y_i/n - \beta_{1}\hspace{0.1em}
\mbox{$\sum_{i=1}^{n}$}\hspace{0.1em} x_{1i}/n - \beta_{2}\hspace{0.1em}
\mbox{$\sum_{i=1}^{n}$}\hspace{0.1em} x_{2i}/n - \beta_{3}\hspace{0.1em}
\mbox{$\sum_{i=1}^{n}$}\hspace{0.1em} x_{3i}/n,\hspace{0.25em} \sigma^2/n \hspace{0.1em}\right)
\nonumber\\[-0.5ex]
\label{equ31}
\end{gather}
\begin{gather}
\beta_2\,|\,\beta_{-2},\sigma^2,y_{+} \sim \mbox{N}\hspace{-0.1em} \left(
\frac{\sum_{i=1}^{n}\hspace{-0.05em} x_{2i}y_{i} - \beta_{0}\sum_{i=1}^{n}\hspace{-0.05em}
x_{2i} - \beta_{1}\sum_{i=1}^{n}\hspace{-0.05em} x_{1i} x_{2i} -
\beta_{3}\sum_{i=1}^{n}\hspace{-0.05em} x_{2i} x_{3i}}{\sum_{i=1}^{n}\hspace{-0.05em}
x_{2i}^{2}},\hspace{0.3em} \frac{\sigma^2}{\sum_{i=1}^{n}\hspace{-0.05em} x_{2i}^{2}}
\right) \nonumber\\[1ex]
\label{equ36}\\[1ex]
\hspace{-0.2em}\sigma^2\,|\,\beta_0,\redots,\beta_3,y_{+} \sim \mbox{Inv-Gamma}\hspace{-0.05em}
\left( \hspace{0.05em}n/2,\hspace{0.25em} \mbox{$\sum_{i=1}^{n}$} (y_i - \beta_0 - \beta_1 x_{1i}
- \beta_2 x_{2i} - \beta_3 x_{3i})^2/2 \hspace{0.05em}\right)
\label{equ37}
\end{gather}
\par \vspace{1ex} \noindent
where $\beta_{-j}$ denotes the set of all the regression coefficients except $\beta_j$.

Now let us provide more details with regard to what was known about the coefficient $\beta_3$
before the data were observed. In particular, let us assume that conditional on all other
parameters in the model being known, the scenario of interest of Section~\ref{sec5} would apply if
the general parameter $\theta_j$ was taken as being $\beta_3$, with the interval $[\theta_{j0},
\theta_{j1}]$ in this scenario now being specified as simply the interval $[-\delta, \delta]$,
where $\delta \geq 0$.
We will therefore construct the full conditional post-data density of $\beta_3$ using the type of
bispatial inference outlined in Section~\ref{sec5}, which implies that, from now on, this density
will be denoted as $b(\beta_3\,|\,\beta_{-3},\sigma^2,y_{+})$.

In particular to do this, the test statistic $T(x)$ as defined in Section~\ref{sec5}, which now
needs to be denoted as $T(y_{+})$, will be assumed to be the least squares estimator of $\beta_3$
under the condition that all other parameters are known, i.e.\ the estimator:
\vspace{1.5ex}
\begin{equation}
\label{equ32}
\bm\hat{\beta}_3 = \frac{\sum_{i=1}^{n}\hspace{-0.05em} x_{3i}y_{i} -
\beta_{0}\sum_{i=1}^{n}\hspace{-0.05em} x_{3i}
- \beta_{1}\sum_{i=1}^{n}\hspace{-0.05em} x_{1i} x_{3i}
- \beta_{2}\sum_{i=1}^{n}\hspace{-0.05em} x_{2i} x_{3i}}{\sum_{i=1}^{n}\hspace{-0.05em} x_{3i}^{2}}
\vspace{1.5ex}
\end{equation}
which is a reasonable assumption to make since, under this condition, it is a sufficient statistic
for $\beta_3$ that satisfies the second criterion given in Section~\ref{sec5} for being the
sta-{\linebreak}tistic $T(y_{+})$.
Observe that this estimator has a sampling distribution that is defined by:
\vspace{0.5ex}
\[
\bm\hat{\beta}_3 \sim \mbox{N}\hspace{-0.1em} \left(\hspace{0.05em} \beta_3,\hspace{0.15em}
\sigma^2/\hspace{0.1em} \mbox{$\sum_{i=1}^{n}$}\hspace{0.1em} x_{3i}^{2} \hspace{0.05em}\right)
\vspace{0.5ex}
\]
Therefore, the hypotheses $H_{P}$ and $H_{S}$ defined in Section~\ref{sec5} that \pagebreak are
applicable in the case where $\bm\hat{\beta}_3 \leq 0$, i.e.\ the hypotheses in
equations~(\ref{equ5}) and~(\ref{equ6}), can now be expressed as:
\begin{gather}
H_{P}: \beta_3 \geq -\delta\nonumber \\[1ex]
H_{S}: \rho (\widehat{B}_3^* < \bm\hat{\beta}_3) \leq \Phi\hspace{-0.05em}\left(
(\bm\hat{\beta}_3 + \delta)(1/\sigma) \sqrt{{\mbox{$\sum_{i=1}^{n}$}\hspace{0.1em}
x_{3i}^{2}}}\hspace{0.1em} \right)\ \
(=\pvalue) \label{equ47}
\end{gather}
\par \vspace{1.5ex} \noindent
where $\Phi(\hspace{0.1em})$ again denotes the standard normal distribution function, while
$\widehat{B}_3^*$ is the estimator $\bm\hat{\beta}_3$ calculated exclusively on the basis of an
as-yet-unobserved sample of $n$ additional data points
$Y_{+}^*=\{(Y_i^*, x_{1i}, x_{2i}, x_{3i}):i=1,2,\ldots,n\}$ generated according to the regression
model in equation~(\ref{equ26}), where the values of the covariates $x_1$, $x_2$ and $x_3$ are
assumed to be the same as in the original sample.
On the other hand, the hypotheses $H_{P}$ and $H_{S}$ that apply if $\bm\hat{\beta}_3 > 0$, i.e.\
the hypotheses in equations~(\ref{equ10}) and~(\ref{equ7}), can now be expressed as:
\vspace{-1ex}
\begin{gather}
H_{P}: \beta_3 \leq \delta\nonumber \\[1ex]
H_{S}: \rho (\widehat{B}_3^* > \bm\hat{\beta}_3) \leq 1 - \Phi\hspace{-0.05em}\left(
(\bm\hat{\beta}_3 - \delta)(1/\sigma) \sqrt{{\mbox{$\sum_{i=1}^{n}$}\hspace{0.1em}
x_{3i}^{2}}}\hspace{0.1em} \right)\ \
(=\pvalue) \label{equ34}
\end{gather}
\par \vspace{1.5ex}
Also, let us assume, quite reasonably, that the fiducial density $f_{S}(\theta_j\,|\,x)$ that is
required by equations~(\ref{equ8}) and~(\ref{equ19}), i.e.\ the density
$f_{S}(\beta_3\,|\,\beta_{-3},\sigma^2,y_{+})$ in the present case, is derived on the basis of the
strong fiducial argument with the fiducial statistic $Q(y_{+})$ specified as being a sufficient
statistic for $\beta_3$, e.g.\ one of the sufficient statistics for $\beta_3$ \linebreak given in
equations~(\ref{equ27}) and~(\ref{equ32}).
Under these assumptions, the fiducial density in question is determined in a similar way to how the
fiducial densities in equations~(\ref{equ31}), (\ref{equ36}) and~(\ref{equ37}) were determined, and
in particular is given by the expression:
\vspace{0.5ex}
\begin{equation}
\label{equ38}
\beta_3\,|\,\beta_{-3},\sigma^2,y_{+} \sim \mbox{N}\hspace{-0.1em} \left( \bm\hat{\beta}_3,
\hspace{0.15em} \sigma^2/ \hspace{0.1em}\mbox{$\sum_{i=1}^{n}$}\hspace{0.1em} x_{3i}^{2} \right)
\vspace{0.5ex}
\end{equation}

On the other hand, it will be assumed that we knew enough \pagebreak about the coefficient
$\beta_1$ before the data were observed such that it is possible to adequately represent our
pre-data knowledge about this coefficient by placing a probability density function over this
coefficient conditional on all other parameters being known, i.e.\ the density
$p(\beta_1\,|\,\beta_{-1},\linebreak \sigma^2)$.
To give an example, let this density function be defined by:
\begin{equation}
\label{equ33}
\beta_1\, |\,\beta_{-1}, \sigma^2 \sim \mbox{N} (\mu_0, \sigma_0^2)
\end{equation}
where $\mu_0$ and $\sigma_0>0$ are given constants.
Treating this choice of the density $p(\beta_1\,|\,\beta_{-1},\linebreak \sigma^2)$ as a prior
density and combining it with the likelihood function in this case, under the Bayesian paradigm,
leads to a full conditional posterior density of $\beta_1$, i.e.\ the density
$p(\beta_1\,|\,\beta_{-1},\sigma^2,y_{+})$, that can be expressed as:
\vspace{1.5ex}
\[
\beta_1\,|\,\beta_{-1},\sigma^2,y_{+} \sim \mbox{N}\hspace{-0.05em} \left( \sigma_1^2 \left[
\frac{\bm\hat{\beta}_1 \sum_{i=1}^n\hspace{-0.05em} x_{1i}^2}{\sigma^2} + \frac{\mu_0}{\sigma_0^2}
\right]\hspace{-0.1em},\hspace{0.25em} \sigma_1^2 \right)
\]
where
\[
\sigma_1^2 = \left( (\mbox{$\sum_{i=1}^n\hspace{-0.05em} x_{1i}^2$} / \sigma^2) +
(1 / \sigma_0^2) \right)^{-1}
\vspace{-1ex}
\]
and
\vspace{1ex}
\[
\bm\hat{\beta}_1 = \frac{\sum_{i=1}^{n}\hspace{-0.05em} x_{1i}y_{i} -
\beta_{0}\sum_{i=1}^{n}\hspace{-0.05em}  x_{1i}
- \beta_{2}\sum_{i=1}^{n}\hspace{-0.05em} x_{1i} x_{2i}
- \beta_{3}\sum_{i=1}^{n}\hspace{-0.05em} x_{1i} x_{3i}}{\sum_{i=1}^{n}\hspace{-0.05em} x_{1i}^{2}}
\vspace{3.5ex}
\]

To illustrate this example, Figure~4 shows some results from running a Gibbs sampler with a uniform
random scanning order of the parameters $\beta_0$, $\beta_1$, $\beta_2$, $\beta_3$ and $\sigma^2$
on the basis of the full conditional post-data densities of these parameters that have just been
detailed, i.e.\ the fiducial densities $f(\beta_0\,|\,\beta_{-0},\sigma^2,y_{+})$,
$f(\beta_2\,|\,\beta_{-2},\sigma^2,y_{+})$ and $f(\sigma^2\,|\,\beta_{0}, \linebreak
\redots,\beta_{3},y_{+})$ defined by equations~(\ref{equ31}), (\ref{equ36}) and~(\ref{equ37}), the
post-data density (derived using bispatial inference) $b(\beta_3\,|\,\beta_{-3},\sigma^2,y_{+})$
and the posterior density $p(\beta_1\,|\,\beta_{-1},\sigma^2,y_{+})$ defined by the equation just
given.
In particular, the histograms in Figures~4(a) to~4(d) represent the distributions of the values of
the coefficients $\beta_1$, $\beta_2$, $\beta_3$ and \pagebreak the standard devia-{\linebreak}tion
$\sigma$, respectively, over a single run of ten million samples of all five model parameters
generated by the Gibbs sampler after allowing for its burn-in phase by discarding a preceding run
of five thousand samples. (For reasons of space, a histogram of the generated values of the
intercept coefficient $\beta_0$ is not given.)
The sampling of the density $b(\beta_3\,|\,\beta_{-3},\sigma^2,y_{+})$ was based on the Metropolis
algorithm, while the sampling of each of the other four full conditional post-data densities was
independent from the preceding iterations.

\begin{figure}[!t]
\begin{center}
\noindent
\makebox[\textwidth]{\includegraphics[width=7.25in]{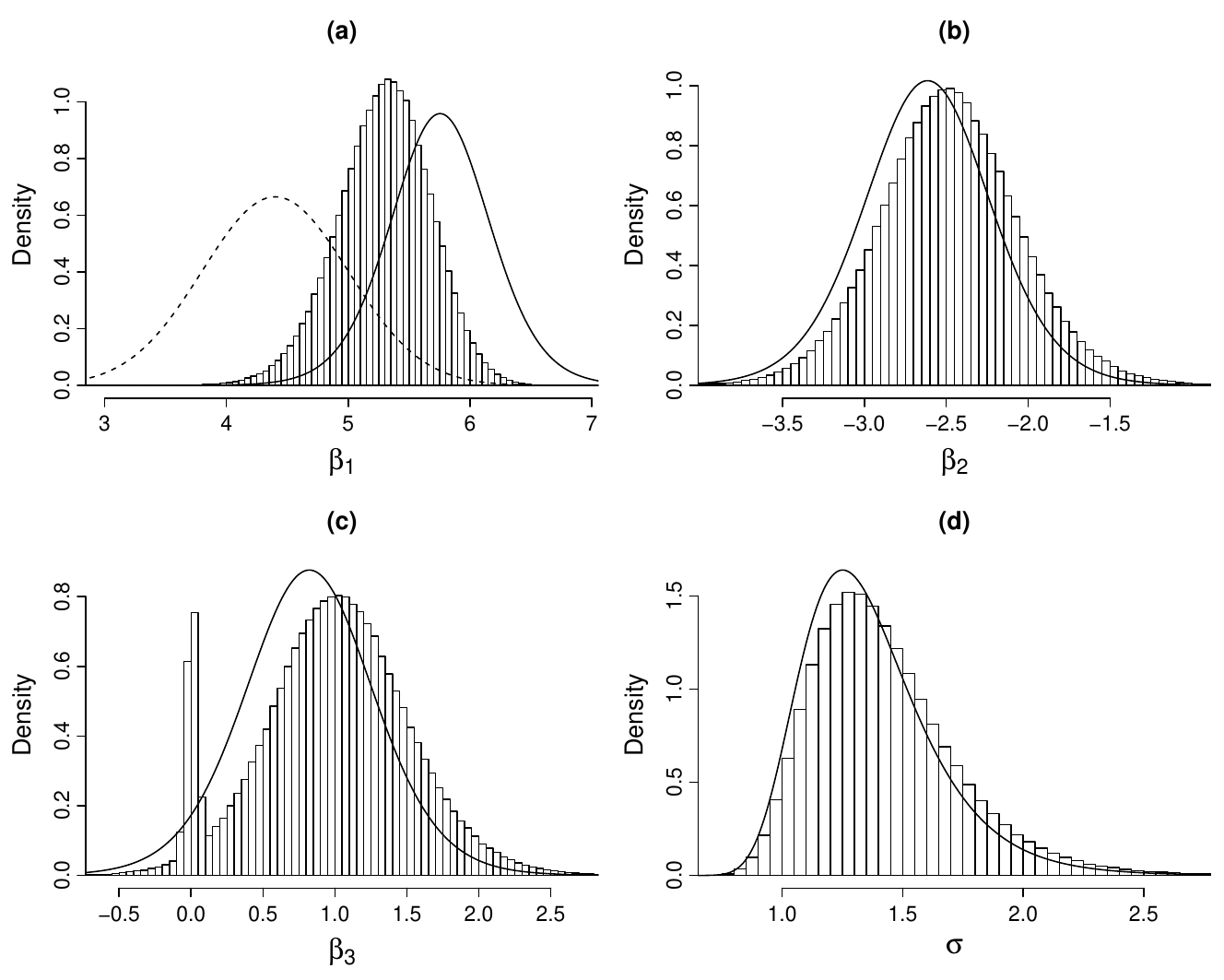}}
\caption{\small{Conditional prior density of one parameter, namely $\beta_1$, and marginal
post-data densities of four parameters $\beta_1$, $\beta_2$, $\beta_3$ and $\sigma$ of a normal
linear regression model}}
\end{center}
\end{figure}

Moreover, the values for the response variable $Y$ in the observed data set $y_{+}$ were a typical
sample of $n=18$ such values generated according to the regression model in equation~(\ref{equ26})
with $\beta_0=0$, $\beta_1=5$, $\beta_2=-2$, $\beta_3=1$ and $\sigma=1.5$, and with the values of
the covariates $x_1$, $x_2$ and $x_3$ in this data set chosen without replacement from the 27
combinations of values for these covariates that are possible if each covariate can only take the
value $-1$, $0$ or $1$.
In particular, the way these covariate values were selected resulted in:
$\sum x_{1i} = -1$, $\sum x_{2i} = 2$, $\sum x_{3i} = 1$, $\sum x_{1i} x_{2i}$ = 3,
$\sum x_{1i} x_{3i} = 4$ and $\sum x_{2i} x_{3i} = -3$.
In addition, the specification of the posterior density $p(\beta_1\,|\,\beta_{-1},\sigma^2,y_{+})$
was completed by setting the constants $\mu_0$ and $\sigma_0$, i.e.\ the constants that control the
choice of the prior density of $\beta_1$ in equation~(\ref{equ33}), to be 4.4 and 0.6 respectively.
On the other hand, with regard to how the post-data density
$b(\beta_3\,|\,\beta_{-3},\sigma^2,y_{+})$ was fully determined, the constant $\delta$ was assumed
to be equal to 0.1, and the probabilities $\kappa$ that would be assigned to the hypothesis $H_{S}$
as defined by either equation~(\ref{equ47}) or equa\-tion~(\ref{equ34}) for different values of all
the model parameters except $\beta_3$ were assumed to be given by the PDO curve with the formula:
$\kappa = \pvalue^{\hspace{0.1em}0.6}$, where, as indicated in equations~(\ref{equ47})
and~(\ref{equ34}), $\pvalue$ is the one-sided P value in whichever definition of the hypothesis
$H_{S}$ is applicable.
Also, in determining the post-data density of $\beta_3$ in question, the density function
$h(\theta_j)$ that appears in equation~(\ref{equ9}), i.e.\ the density $h(\beta_3)$ in the present
case, was defined similar to how a density function of this type was specified in
Section~\ref{sec12}, \linebreak that is, by the expression $\beta_3 \sim \mbox{Beta}\hspace{0.1em}
(4,4,-0.1,0.1)$, where the notation here is the same as used in equation~(\ref{equ35}).

Supplementary to this analysis, there was no suggestion from applying appropriate diagnostics to
multiple runs of the Gibbs sampler from different starting points that it did not have a limiting
distribution.
Furthermore, the Gibbs sampling algorithm was run separately with various very distinct fixed
scanning orders of the five model parameters $\beta_0$, $\beta_1$, $\beta_2$, $\beta_3$ and
$\sigma^2$ in accordance with how a single transition of such an algorithm with a fixed scanning
order was defined in Section~\ref{sec2}.
In doing this, no statistically significant difference was found between the samples of parameter
values aggregated over the runs of the sampler, after excluding the burn-in phase of the sampler,
in using each of the scanning orders concerned, e.g.\ between the various correlation matrices of
the parameters and between the various distributions of each individual parameter, even when the
runs in question were long.
Therefore, on the grounds of what was discussed in Section~\ref{sec2}, it would be reasonable to
conclude that the full conditional densities of the limiting distribution of the original
random-scan Gibbs sampler should be, at the very least, close approximations to the full
conditional densities on which the sampler is based,
i.e.\ the fiducial densities $f(\beta_0\,|\,\beta_{-0},\sigma^2,y_{+})$,
$f(\beta_2\,|\,\beta_{-2},\sigma^2,y_{+})$ and $f(\sigma^2\,|\,\beta_{0},\redots,\beta_{3},y_{+})$,
the post-data density $b(\beta_3\,|\,\beta_{-3},\sigma^2,y_{+})$ and the posterior density
$p(\beta_1\,|\,\beta_{-1},\sigma^2,y_{+})$.

The solid curves overlaid on the histograms in Figures~4(a) to 4(d) are plots of the marginal
densities of the coefficients $\beta_1$, $\beta_2$, $\beta_3$ and the standard deviation $\sigma$,
respectively, over the joint fiducial density of all the parameters in the model that is defined
directly and uniquely by the set of compatible full conditional densities that consists of the
fiducial densities $f(\beta_0\,|\,\beta_{-0},\sigma^2,y_{+})$,
$f(\beta_2\,|\,\beta_{-2},\sigma^2,y_{+})$ and $f(\sigma^2\,|\,\beta_{0},\redots,\beta_{3},y_{+})$
just referred to, which of course are given by equations~(\ref{equ31}), (\ref{equ36})
and~(\ref{equ37}), the fiducial density $f_{S}(\beta_3\,|\,\beta_{-3},\sigma^2,y_{+})$ given by
equation~(\ref{equ38}), and the fiducial density for $\beta_1$ conditional on $\beta_0$, $\beta_2$,
$\beta_3$ and $\sigma^2$ that results from making assumptions that are analogous to those on which
the aforementioned full conditional fiducial densities for $\beta_0$, $\beta_2$ and $\beta_3$ are
based. On the other hand, the dashed curve overlaid on the histogram in Figure~4(a) is a plot of
the conditional prior density of $\beta_1$ given in equation~(\ref{equ33}).

By comparing the histograms in Figures~4(a) to 4(d) with the curves overlaid on them, it can be
seen that the forms of the marginal post-data densities of $\beta_1$, $\beta_2$, $\beta_3$ and
$\sigma$ that are represented by these histograms are consistent with what could have been
intuitively expected given the pre-data beliefs about all of the model parameters that were taken
into account as part of the method of inference that has been described in the present section.

\vspace{3ex}
\subsection{Inference about a bivariate normal distribution}
\label{sec7}

To give a final detailed example of the application of integrated organic inference, let us
consider the problem of making inferences about all five parameters of a bivariate normal density
function, i.e.\ the means $\mu_x$ and $\mu_y$ and the variances $\sigma^2_x$ and $\sigma^2_y$,
respectively, of the two random variables concerned $X$ and $Y$, and the correlation $\tau$ of $X$
and $Y$, on the basis of a sample from this type of density function, i.e.\ the sample
${\tt z}=\{(x_i,y_i) : i=1,2,\ldots,n\}$, where $x_i$ and $y_i$ are the $i$th realisations of $X$
and $Y$ respectively.

In Bowater~(2018a), as a way of addressing this problem, full conditional fiducial densities were
derived either exactly or approximately for each of the parameters $\mu_x$, $\mu_y$, $\sigma^2_x$,
$\sigma^2_y$ and $\tau$ by using appropriately chosen fiducial statistics under the strong fiducial
argument, and then it was illustrated how, on the basis of these conditional densities, what can be
regarded as being a suitable joint fiducial density of these parameters can be obtained by using
the Gibbs sampler within the type of framework outlined in Section~\ref{sec2} of the current paper.
However, for the same kind of reason that was given in relation to the use of a similar method of
inference in the previous section, this particular method is not going to be directly applicable to
the case that will be presently considered.
This is more specifically due to the fact that, although we will assume that nothing or very little
was known about the means $\mu_x$ and $\mu_y$ before the data were observed, by contrast we are
going to assume that there was a substantial amount of pre-data knowledge about the variances
$\sigma^2_x$ and $\sigma^2_y$ and the correlation coefficient $\tau$. To begin with though, let us
clarify how the full conditional post-data densities of $\mu_x$ and $\mu_y$ will be constructed.

In this regard, observe that sufficient statistics for the parameters $\mu_x$ and $\mu_y$
condi-{\linebreak}tional on all parameters except the parameter itself being known are:
\[
q_{x} = \bar{x} - \tau (\sigma_x / \sigma_y)\bar{y}\ \ \ \mbox{and}\ \ \
q_{y} = \bar{y} - \tau (\sigma_y / \sigma_x)\bar{x},
\vspace{0.5ex}
\]
respectively, where $\bar{x} = \sum_{i=1}^{n} x_i$ and $\bar{y} = \sum_{i=1}^{n} y_i$. Therefore,
these two statistics $q_{x}$ and $q_{y}$ will be assumed to be the fiducial statistics $Q({\tt z})$
that will be used in making in\-ferences about $\mu_x$ and $\mu_y$ respectively.
Under this assumption, if $\mu_x$ is the only unknown parameter in the model, then
equation~(\ref{equ3}) will now have the form $q_{x} = \varphi(\Gamma,\mu_x)$, and more specifically
can be expressed as:
\vspace{1.5ex}
\[
\bar{x} - \tau \hspace{-0.05em} \left( \frac{\sigma_x}{\sigma_y} \right) \bar{y} = \mu_x -
\tau \hspace{-0.05em} \left( \frac{\sigma_x}{\sigma_y} \right) \mu_y + \Gamma\hspace{0.15em}
\sqrt{\frac{\sigma_x^2(1-\tau^2)}{n}}
\vspace{2ex}
\]
where the primary r.v.\ $\Gamma \sim \mbox{N}(0,1)$.
Also, given what has been assumed in relation to our pre-data knowledge about $\mu_x$, it would be
quite natural to specify the GPD function for $\mu_x$ as follows: $\omega_G(\mu_x) = a$ for
$\mu_x \in (-\infty,\infty)$, where $a>0$.
This implies that the full conditional fiducial density of $\mu_x$ is defined by:
\vspace{1.5ex}
\begin{equation}
\label{equ43}
\mu_x\, |\,\mu_y,\sigma_x^2,\sigma_y^2,\tau, {\tt z} \sim \mbox{N} \left(\hspace{0.1em} \bar{x} +
\tau\hspace{-0.05em} \left(\frac{\sigma_x}{\sigma_y} \right)\hspace{-0.1em}
(\mu_y - \bar{y}),\hspace{0.4em} \frac{\sigma_x^2(1-\tau^2)}{n}\hspace{0.1em} \right)
\vspace{1.5ex}
\end{equation}
Furthermore, due to the symmetrical nature of the bivariate normal distribution, it should be clear
that, using a GPD function for $\mu_y$ of the same type as just used for $\mu_x$, the full
conditional fiducial density of $\mu_y$ would be defined by:
\vspace{1.5ex}
\begin{equation}
\label{equ44}
\mu_y\, |\,\mu_x,\sigma_x^2,\sigma_y^2,\tau, {\tt z} \sim \mbox{N} \left(\hspace{0.1em} \bar{y} +
\tau\hspace{-0.05em} \left(\frac{\sigma_y}{\sigma_x} \right)\hspace{-0.1em}
(\mu_x - \bar{x}),\hspace{0.4em} \frac{\sigma_y^2(1-\tau^2)}{n}\hspace{0.1em} \right)
\vspace{1.5ex}
\end{equation}

With regard to what was known about the variances $\sigma_x^2$ and $\sigma_y^2$ before the data
were observed, we will assume that it is possible to adequately \pagebreak represent such knowledge
by placing a probability density function over each of these parameters conditional on all
parameters except the parameter itself being known, i.e.\ the densities
$p(\sigma_x^2\,|\,\mu_x, \mu_y,\linebreak \sigma_y^2, \tau)$ and
$p(\sigma_y^2\,|\,\mu_x, \mu_y, \sigma_x^2, \tau)$ respectively.
To give an example, let these density func\-tions for $\sigma_x^2$ and $\sigma_y^2$ be defined
respectively by:
\vspace{0.5ex}
\begin{equation}
\label{equ39}
\sigma_x^2 \sim \mbox{Inv-Gamma}\, (\alpha_x, \beta_x)\ \ \mbox{and}\ \
\sigma_y^2 \sim \mbox{Inv-Gamma}\, (\alpha_y, \beta_y)
\end{equation}
where $\alpha_x$, $\beta_x$, $\alpha_y$ and $\beta_y$ are given positive constants.

Notice that, for the case being considered, the likelihood functions that would be placed over each
of the parameters $\sigma_x^2$ and $\sigma_y^2$ assuming that all parameters except the parameter
itself are known are given by the expressions:
\vspace{1.5ex}
\begin{equation}
\label{equ40}
\hspace{-0.3em}L(\sigma_x^2\,|\,\mu_x,\mu_y,\sigma_y^2,\tau,{\tt z}) = (1/\sigma_x)^n \exp\left(
\frac{-1}{2(1-\tau^2)}\left( \frac{\sum (x'_i)^2}{\sigma_x^2} \right) + \frac{\tau}{1-\tau^2}
\left(\frac{\sum x'_i y'_i}{\sigma_x \sigma_y} \right) \right)
\end{equation}
and
\vspace{-0.5ex}
\begin{equation}
\label{equ41}
\hspace{-0.3em}L(\sigma_y^2\,|\,\mu_x,\mu_y,\sigma_x^2,\tau,{\tt z}) = (1/\sigma_y)^n \exp\left(
\frac{-1}{2(1-\tau^2)}\left( \frac{\sum (y'_i)^2}{\sigma_y^2} \right) + \frac{\tau}{1-\tau^2}
\left( \frac{\sum x'_i y'_i}{\sigma_x \sigma_y} \right) \right)
\vspace{2.5ex}
\end{equation}
respectively, where $x'_{i} = x_{i} - \mu_x$ and $y'_{i} = y_{i} - \mu_y$.
Therefore, if the choices of the densities $p(\sigma_x^2\,|\,\mu_x, \mu_y, \sigma_y^2, \tau)$ and
$p(\sigma_y^2\,|\,\mu_x, \mu_y, \sigma_x^2, \tau)$ in equation~(\ref{equ39}) are treated as prior
densities, it can easily be seen how, by combining these prior densities with the likelihood
functions in equations~(\ref{equ40}) and~(\ref{equ41}) under the Bayesian paradigm, the full
conditional posterior densities of $\sigma_x^2$ and $\sigma_y^2$ can be numerically computed, i.e.\
the posterior densities $p(\sigma_x^2\,|\,\mu_x,\mu_y,\sigma_y^2,\tau,{\tt z})$ and
$p(\sigma_y^2\,|\,\mu_x,\mu_y,\sigma_x^2,\tau,{\tt z})$.

On the other hand, with regard to the beliefs that were held about the correlation coefficient
$\tau$ before the data were observed, let us assume that conditional on all other parameters being
known, the scenario of interest of Section~\ref{sec5} would apply if the general parameter
$\theta_j$ was taken as being $\tau$, with the interval $[\theta_{j0}, \theta_{j1}]$ in
\pagebreak this scenario now being specified as the interval $[-\varepsilon, \varepsilon]$, where
$\varepsilon \geq 0$. As a result, we will now discuss how the full conditional post-data density
of $\tau$ will be constructed by using the type of bispatial inference outlined in
Section~\ref{sec5}, which implies that it will be denoted as the density
$b(\tau\,|\,\mu_x,\mu_y,\sigma_x^2,\sigma_y^2,{\tt z})$.

In this respect, let us begin by pointing out that since, if all parameters except $\tau$ are
known, there exists no sufficient set of univariate statistics for $\tau$ that contains only one
statistic that is not an ancillary statistic, it would seem reasonable to assume that the test
statistic $T({\tt z})$, as generally defined in Section~\ref{sec5}, is the maximum likelihood
estimator of $\tau$ given that all other parameters are known.
It can be shown that this maximum likelihood estimator is the value $\bm\hat{\tau}$ that solves the
following cubic equation:
\vspace{2ex}
\[
-n\bm\hat{\tau}^{\hspace{0.05em}3} + \left( \frac{\sum_{i=1}^n x'_i y'_i}{\sigma_x \sigma_y}
\right)\hspace{-0.1em} \bm\hat{\tau}^{\hspace{0.05em}2} + \left( n -
\frac{\sum_{i=1}^n (x'_i)^2}{\sigma_x^2} - \frac{\sum_{i=1}^n (y'_i)^2}{\sigma_y^2} \right)
\hspace{-0.1em} \bm\hat{\tau} \hspace{0.05em} + \hspace{0.03em}
\frac{\sum_{i=1}^n x'_i y'_i}{\sigma_x \sigma_y} = 0
\vspace{2ex}
\]

Now, it is well known that a maximum likelihood estimator of a parameter is usually asymptotically
normally distributed with mean equal to the true value of the parameter, and variance equal to the
inverse of the Fisher information with respect to that parameter. (To clarify, this is the Fisher
information obtained via differentiating the logarithm of the likelihood function with respect to
the parameter concerned.) For this reason, if $n$ is large, the sampling density function of the
maximum likelihood estimator $\bm\hat{\tau}$ just defined can be approximately expressed as
follows:
\vspace{-0.25ex}
\begin{equation}
\label{equ51}
\bm\hat{\tau} \sim \mbox{N} ( \tau, 1/ \mathcal{I}(\tau) )
\vspace{-0.25ex}
\end{equation}
where $\mathcal{I}(\tau)$ is the Fisher information of the likelihood function in this example
with respect to $\tau$ assuming all other parameters are known, which is in fact given by:
\vspace{1.5ex}
\[
\mathcal{I}(\tau) = \frac{n(1+\tau^2)}{(1-\tau^2)^2}
\pagebreak
\]

Using this approximation, the hypotheses $H_{P}$ and $H_{S}$ defined in Section~\ref{sec5} that are
applicable in the case where $\bm\hat{\tau} \leq 0$, i.e.\ the hypotheses in equations~(\ref{equ5})
and~(\ref{equ6}), can now be expressed as:
\vspace{-1.5ex}
\begin{gather}
H_{P}: \tau \geq -\varepsilon \label{equ48}\\[0.5ex]
H_{S}: \rho (\widehat{T}^* < \bm\hat{\tau}) \leq \Phi\hspace{-0.05em}\left(
(\bm\hat{\tau} + \varepsilon)\sqrt{\mathcal{I}(\varepsilon)}\hspace{0.1em} \right)\ \
(=\pvalue) \label{equ49}
\end{gather}
\par \vspace{1ex} \noindent
where $\widehat{T}^*$ is the estimator $\bm\hat{\tau}$ calculated exclusively on the basis of an
as-yet-unobserved sample of $n$ additional data points $\{(X_i^*, Y_i^*):i=1,2,\ldots,n\}$ drawn
from the bivariate normal density function being studied, and $\Phi(\hspace{0.1em})$ is again the
standard normal distribu\-tion function.
On the other hand, the hypotheses $H_{P}$ and $H_{S}$ that apply if $\bm\hat{\tau} > 0$, i.e.\
\linebreak the hypotheses in equations~(\ref{equ10}) and~(\ref{equ7}), can now be expressed as:
\begin{gather}
H_{P}: \tau \leq \varepsilon \label{equ50}\\[0.5ex]
H_{S}: \rho (\widehat{T}^* > \bm\hat{\tau}) \leq 1 - \Phi\hspace{-0.05em}\left(
(\bm\hat{\tau} - \varepsilon)\sqrt{\mathcal{I}(\varepsilon)} \hspace{0.1em}\right)\ \
(=\pvalue) \label{equ45}
\end{gather}
\par \vspace{1ex} \noindent
We should point out that if the estimator $\bm\hat{\tau}$ did indeed have the normal distribution
given in equation~(\ref{equ51}), then it can be easily shown that this estimator would satisfy the
second criterion given in Section~\ref{sec5} for being a valid test statistic $T({\tt z})$, which
would in turn imply that the hypotheses $H_{P}$ and $H_{S}$ as defined in equations~(\ref{equ48})
and~(\ref{equ49}) would be equivalent, and also that these hypotheses as defined in
equations~(\ref{equ50}) and~(\ref{equ45}) \linebreak would be equivalent.

To determine the fiducial density $f_{S}(\theta_j\,|\,x)$ that is required by
equations~(\ref{equ8}) and~(\ref{equ19}), i.e.\ the density
$f_S(\tau\,|\,\mu_x, \mu_y, \sigma_x^2, \sigma_y^2,{\tt z})$ in the present case, let us begin by
assuming that the maximum likelihood estimator $\bm\hat{\tau}$ is the fiducial statistic
$Q({\tt z})$, which is actually the \linebreak choice that was made for this statistic $Q({\tt z})$
in the aforementioned example in Bowater~(2018a) when fiducial inference was used in this type of
situation, \pagebreak i.e.\ in the situation where $\tau$ is the only unknown parameter.
However, instead of assuming that the sampling density function of $\bm\hat{\tau}$ is a normal
density as has just been done, and as was done in the context of current interest in
Bowater~(2018a), let us assume that it is a transformation of $\bm\hat{\tau}$ that is normally
distributed, namely the function $\tanh^{-1} (\bm\hat{\tau})$.
The reason for doing this is that it can be shown that, under this latter assumption, a generally
better approximation to the sampling density of $\bm\hat{\tau}$ can be obtained than under the
former assumption, except, that is, when $\tau$ is close to zero. Notice that this exception is the
reason why this alternative assumption was not the preferred assumption in the preceding discussion
in order to derive approximate forms of the hypothesis $H_{S}$.
More specifically, it will be assumed that the density function of $\tanh^{-1}(\bm\hat{\tau})$ is
directly specified (and the density function of $\bm\hat{\tau}$ is therefore indirectly specified)
by the expression:
\vspace{0.5ex}
\[
\tanh^{-1}(\bm\hat{\tau}) \sim \mbox{N}\hspace{0.05em} ( \tanh^{-1}(\tau),\hspace{0.05em} 1/
\mathcal{I}(\tanh^{-1}\tau) )
\vspace{0.5ex}
\]
where $\mathcal{I}(\tanh^{-1} \tau)$ is the Fisher information with respect to the quantity
$\tanh^{-1}(\tau)$ assuming all parameters except $\tau$ are known, which is in fact given by:
\vspace{0.5ex}
\[
\mathcal{I}(\tanh^{-1} \tau) = n(1+\tau^2)
\vspace{0.5ex}
\]

Allowing $\tanh^{-1}(\bm\hat{\tau}$) to take the role of the statistic $Q({\tt z})$, and using the
approxima-{\linebreak}tion to the density function of this statistic $\tanh^{-1}(\bm\hat{\tau}$)
just given, we can therefore approximate equation~(\ref{equ3}) in the case where $\tau$ is the only
unknown \vspace{1.5ex} parameter as follows:
\begin{equation}
\label{equ42}
\tanh^{-1}(\bm\hat{\tau}) = \varphi(\Gamma,\tau) = \tanh^{-1}(\tau) +
\frac{\Gamma}{\sqrt{n(1+\tau^2)}}
\vspace{2.5ex}
\end{equation}
where the primary r.v.\ $\Gamma \sim \mbox{N}(0,1)$.
Although it can be shown that this equation does not generally satisfy Condition~1 of
Section~\ref{sec4}, it is the case, on the other hand, that if $\Gamma$ is generated from a
standard normal density function truncated \pagebreak to lie in a given interval
$(-\mathtt{v},\mathtt{v})$ where $\mathtt{v}>0$, then this condition will be satisfied for very
large values of $\mathtt{v}$ under the restriction that $n$ is not too small and $\bm\hat{\tau}$ is
not very close to $-1$ or $1$. For example, if $n=100$ and $|\bm\hat{\tau}| < 0.999$, then
Condition~1 will be satisfied not only for small values of $\mathtt{v}$, but even if $\mathtt{v}$
is chosen to be as high as 36, and will be satisfied for \linebreak substantially larger values of
$\mathtt{v}$ as $|\bm\hat{\tau}|$ becomes smaller.

We will therefore make use of equation~(\ref{equ42}) under the assumption that the primary r.v.\
$\Gamma$ follows the truncated normal density function just mentioned with $\mathtt{v}$ chosen to
be equal to or not far below the largest possible value of $\mathtt{v}$ that is consistent with
equation~(\ref{equ42}) satisfying Condition~1.
Also, since the fiducial density $f_S(\tau\,|\,\mu_x, \mu_y, \sigma_x^2, \sigma_y^2, {\tt z})$
needs to be derived under the assumption that, given the values of the conditioning parameters
$\mu_x$, $\mu_y$, $\sigma_x^2$ and $\sigma_y^2$, there would have been no or very little pre-data
knowledge about $\tau$, it will be quite naturally assumed that the GPD function of $\tau$ is
specified as follows: $\omega_G(\tau) = b$ if $-1 \leq \tau \leq 1$ and 0 otherwise, where $b>0$.
Under the assumptions that have just been made, applying the principle outlined in
Section~\ref{sec4} for deriving a fiducial density of the general type
$f(\theta_j\,|\,\theta_{-j},x)$, i.e.\ Principle~1 of Bowater~(2019a), leads to an approximation to
the full conditional fiducial density of $\tau$ that is given by:
\vspace{2ex}
\[
f_{S}( \tau\,|\,\mu_x,\mu_y,\sigma_x^2,\sigma_y^2,{\tt z}) = \psi_t(\gamma) \left|
\frac{d\gamma}{d\tau} \right|\ \ \ \mbox{if $\tau \in (\tau_0, \tau_1)$ and is zero otherwise}
\vspace{2ex}
\]
where $\gamma$ is the value of $\Gamma$ that solves equation~(\ref{equ42}) for the given value of
$\tau$, i.e.\
\vspace{0.5ex}
\[
\gamma = (\tanh^{-1}(\bm\hat{\tau}) - \tanh^{-1}(\tau))\hspace{0.05em}n^{0.5}(1+\tau^2)^{0.5}
\vspace{0.5ex}
\]
while $\psi_t(\gamma)$ is the standard normal density function truncated to lie in the interval
\linebreak $(-\mathtt{v},\mathtt{v})$ evaluated at $\gamma$, and finally $(\tau_0, \tau_1)$ is the
interval of values of $\tau$ that, according to equation~(\ref{equ42}), correspond to $\gamma$
lying in the interval $(-\mathtt{v},\mathtt{v})$. With the assumption having been made that the
fiducial density $f_{S}( \tau\,|\,\mu_x,\mu_y,\sigma_x^2,\sigma_y^2,{\tt z})$ \pagebreak is
approximately determined in this manner, it can be easily seen how the specification of the
post-data density $b(\tau\,|\,\mu_x,\mu_y,\sigma_x^2,\sigma_y^2,{\tt z})$ can be completed by using
the criteria of Section~\ref{sec5}. 

To illustrate this example, Figure~5 shows some results from running a Gibbs sampler with a uniform
random scanning order of the parameters $\mu_x$, $\mu_y$, $\sigma_x^2$, $\sigma_y^2$ and $\tau$ on
the basis of the full conditional post-data densities of these parameters that have just been
detailed, i.e.\ the fiducial densities
$f(\mu_x\, |\,\mu_y,\sigma_x^2,\sigma_y^2,\tau,{\tt z})$ and
$f(\mu_y\, |\,\mu_x,\sigma_x^2,\sigma_y^2,\tau,{\tt z})$ defined by equations~(\ref{equ43})
and~(\ref{equ44}), the posterior densities (derived using Bayesian inference)
$p(\sigma_x^2\,|\,\mu_x,\mu_y,\sigma_y^2,\tau,{\tt z})$ and
$p(\sigma_y^2\,|\,\mu_x,\mu_y,\sigma_x^2,\tau,{\tt z})$ and the post-data density (derived using
bispatial inference) $b(\tau\,|\,\mu_x,\mu_y,\sigma_x^2,\sigma_y^2,{\tt z})$.
In particular, the histograms in Figures~5(a) to~5(e) represent the distributions of the values of
$\mu_x$, $\mu_y$, $\sigma_x$, $\sigma_y$ and $\tau$, respectively, over a single run of ten million
samples of these parameters generated by the Gibbs sampler after allowing for its burn-in phase by
discarding a preceding run of five thousand samples.
The sampling of each of the densities
$p(\sigma_x^2\,|\,\mu_x,\mu_y,\sigma_y^2,\tau,{\tt z})$, \linebreak
$p(\sigma_y^2\,|\,\mu_x,\mu_y,\sigma_x^2,\tau,{\tt z})$ and
$b(\tau\,|\,\mu_x,\mu_y,\sigma_x^2,\sigma_y^2,{\tt z})$ was based on the Metropolis algorithm,
while the sampling of each of the densities $f(\mu_x\,|\,\mu_y,\sigma_x^2,\sigma_y^2,\tau,{\tt z})$
and $f(\mu_y\,|\,\mu_x,\sigma_x^2,\sigma_y^2,\tau,{\tt z})$ was independent from the preceding
iterations.

\begin{figure}[!p]
\begin{center}
\noindent
\makebox[\textwidth]{\includegraphics[width=7in]{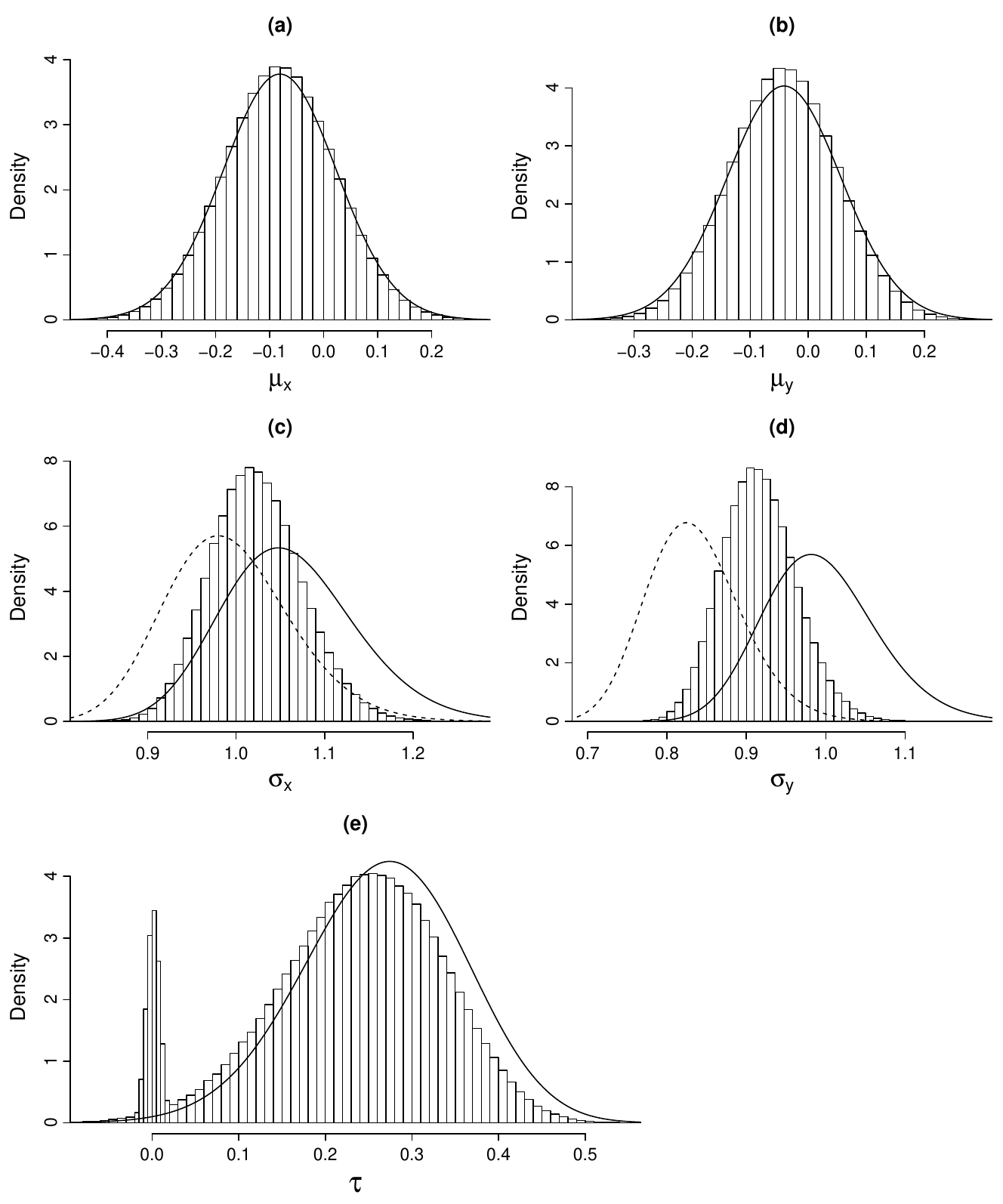}}
\caption{\small{Conditional prior densities of two parameters, namely $\sigma_x$ and $\sigma_y$,
and marginal post-data densities of all five parameters of a bivariate normal distribution}}
\end{center}
\end{figure}

Moreover, the observed data set ${\tt z}$ was a typical sample of $n=100$ data points from a
bivariate normal distribution with $\mu_x=0$, $\mu_y=0$, $\sigma_x=1$, $\sigma_y=1$ and $\tau=0.3$.
In addition, the specification of the posterior densities
$p(\sigma_x^2\,|\,\mu_x,\mu_y,\sigma_y^2,\tau,{\tt z})$ and
$p(\sigma_y^2\,|\,\mu_x, \linebreak \mu_y,\sigma_x^2,\tau,{\tt z})$ were completed by assuming the
values of the constants $\alpha_x$, $\beta_x$, $\alpha_y$ and $\beta_y$, i.e.\ the constants that
control the choice of the prior densities of $\sigma_x^2$ and $\sigma_y^2$ in
equation~(\ref{equ39}), were set as follows: $\alpha_x=49.5$, $\beta_x=48$, $\alpha_y=49.5$ and
$\beta_y=34$.
On the other hand, with regard to how the post-data density
$b(\tau\,|\,\mu_x,\mu_y,\sigma_x^2,\sigma_y^2,{\tt z})$ was fully determined, the constant
$\varepsilon$ was assumed to be equal to 0.02, and the probabilities $\kappa$ that would be
assigned to the hypotheses $H_{S}$ in equations~(\ref{equ49}) and (\ref{equ45}) for different
values of all the parameters except $\tau$ were assumed to be given by the PDO curve with, once
more, the formula: $\kappa = \pvalue^{\hspace{0.1em}0.6}$, where, as indicated in these earlier
equations, $\pvalue$ is the one-sided P value in the definition of the hypothesis $H_{S}$ that is
applicable. Also, in determining the post-data density of $\tau$ in question, the density function
$h(\theta_j)$ that appears in equation~(\ref{equ9}), i.e.\ the density $h(\tau)$ in the present
case, was defined similar to how a density function of this type was specified in earlier examples,
that is, by the expression $\tau \sim \mbox{Beta}\hspace{0.1em} (4,4,-0.02,0.02)$, where the
notation here is again as used in \linebreak equation~(\ref{equ35}).

Supplementary to this analysis, there was no suggestion from applying appropriate diagnostics to
multiple runs of the Gibbs sampler from different starting points that it did not have a limiting
distribution.
Furthermore, after excluding the burn-in phase of the sampler, no statistically significant
difference was found between the samples of parameter values aggregated over the runs of the
sampler in using various very distinct fixed scanning orders of the five model parameters $\mu_x$,
$\mu_y$, $\sigma_x^2$, $\sigma_y^2$ and $\tau$, with a single transition of the sampler defined in
the same way as in previous examples, even when the runs in question were long.
Taking into account what was discussed in \linebreak Section~\ref{sec2}, we can reasonably
conclude, therefore, that the full conditional densities of the limiting distribution of the
original random-scan Gibbs sampler should be, at the very least, close approximations to the full
conditional densities on which the sampler is based, i.e.\ the fiducial densities
$f(\mu_x\, |\,\mu_y,\sigma_x^2,\sigma_y^2,\tau,{\tt z})$ and
$f(\mu_y\, |\,\mu_x,\sigma_x^2,\sigma_y^2,\tau,{\tt z})$,
the posterior densities $p(\sigma_x^2\,|\,\mu_x,\mu_y,\sigma_y^2,\tau,{\tt z})$ and
$p(\sigma_y^2\,|\,\mu_x,\mu_y,\sigma_x^2,\tau,{\tt z})$ and the post-data density
$b(\tau\,|\,\mu_x,\mu_y,\sigma_x^2,\sigma_y^2,{\tt z})$.

The solid curves overlaid on the histograms in Figures~5(a) and 5(c) are plots of the marginal
fiducial densities of the parameters $\mu$ and $\sigma$, respectively, as defined by
equations~(\ref{equ16}) and~(\ref{equ15}) that would apply if the data set of interest only
consisted of the observed values of the variable $X$, i.e.\ $\{x_i:i=1,2,\ldots,100\}$, while in
Figures~5(b) and~5(d), the solid curves represent, respectively, the marginal fiducial densities of
$\mu$ and $\sigma$ defined in the same way except that these densities correspond to treating the
observed values of the variable $Y$ rather than the variable $X$, i.e.\ the set of values
$\{y_i:i=1,2,\ldots,100\}$, as being the data set $x$ in the equations being discussed.
On the other hand, the dashed curves overlaid on the histograms in Figures~5(c) and~5(d) are plots
of the conditional prior densities for $\sigma_x$ and $\sigma_y$, respectively, as defined in
equation~(\ref{equ39}).

Finally, the solid curve overlaid on the histogram in Figure~5(e) is a plot of a confidence density
function for the parameter $\tau$. In general, a density function of this type corresponds to a set
of confidence intervals that have a varying coverage probability for the parameter concerned, see
for example Efron~(1993) for further clarification.
More specifically, for the plot being considered, these confidence intervals for $\tau$ were
constructed on the basis of summarising the data set ${\tt z}$ by the sample correlation
coefficient $\mathtt{r}$, and then assuming that the Fisher transformation of this coefficient,
i.e.\ the transformation $\tanh^{-1}(\mathtt{r})$, has a normal sampling distribution with mean
$\tanh^{-1}(\tau)$ and variance $1/(n-3)$, which is a standard method that is used in practice to
form confidence intervals for the correlation $\tau$.

Similar to earlier examples, it can be seen from comparing the histograms in Figures~5(a) to~5(d)
with the curves overlaid on them that the forms of the marginal post-data densities of $\mu_x$
$\mu_y$, $\sigma_x$ and $\sigma_y$ that are represented by these histograms are consistent with
what we would have intuitively expected given the pre-data beliefs about these parameters and the
correlation $\tau$ that have been taken into account.
Furthermore, we can observe that the marginal post-data density for $\tau$ represented by the
histogram in Figure~5(e) differs substantially from the curve overlaid on this histogram, i.e.\ the
aforementioned type of confidence density function for $\tau$, particularly with regard to the
amount of probability mass that these two density functions assign to values of\hspace{0.03em}
$\tau$ close to zero.
This arguably gives an indication of how inadequate it would be, in this example, to attempt to
make inferences about the correlation $\tau$ using the standard type of confidence intervals for
$\tau$ on which the overlaid curve in question is based.

\vspace{3ex}
\subsection{Summary of other examples}

As part of the discussion of the examples that were outlined in the preceding sections, reference
was made to additional examples from Bowater~(2018a), Bowater~(2019a) and Bowater~(2019b) that fit
within the inferential framework that has been put forward in the present paper. Here the
opportunity will be taken to highlight examples of a similar kind from these earlier papers that
have not been mentioned up to this point.

To begin with, let us remark that in Bowater~(2019a), organic fiducial inference was applied to the
problem of making post-data inferences about discrete probability distributions that naturally only
have one unknown parameter, in particular the binomial and Poisson distributions, and as a result,
a fiducial density for the parameter concerned was determined.
With regard to making inferences about a binomial proportion, the application of the method of
inference in question represents, of course, a special case of the type of scenario discussed in
Section~\ref{sec11}, i.e.\ the case where the population proportion $\pi_1$ in this latter example
is set to zero.
Furthermore, the problem of making post-data inferences about a binomial proportion was addressed
in Bowater~(2019b) by using the type of bispatial inference that was described in
Section~\ref{sec5}.

On the other hand, in Bowater~(2018a), joint post-data densities for the two parameters of the
Pareto, gamma and beta distributions were determined by using the type of framework that was
outlined in Section~\ref{sec2} on the basis of full conditional post-data densities of the
parameters concerned that were formed by applying, in effect, organic fiducial inference, i.e.\ all
these full conditional and joint post-data densities were, in fact, fiducial densities.
In addition, the post-data density for a relative risk $\pi_t/\pi_c$ was determined in
Bowater~(2019b) by using the kind of framework of Section~\ref{sec2} on the basis of full
conditional post-data densities for the binomial proportions $\pi_t$ and $\pi_c$ that were formed
by applying the type of bispatial inference detailed in Section~\ref{sec5} in a way that meant that
dependence would, in general, exist between $\pi_t$ and $\pi_c$ in the joint post-data density of
these parameters.
Finally, in Bowater~(2018a), a method that was, in effect, organic fiducial inference was applied
to the problem of making post-data inferences about the difference between the means of two normal
density functions that have unknown variances on the basis of independent samples from the two
density functions concerned, i.e.\ the Behrens\hspace{0.05em}-Fisher problem.

\vspace{3ex}
\section{Defence and discussion of the theory}
\label{sec8}

There now follows a discussion of the theory put forward in the present paper, i.e.\ integrated
organic inference, arranged as a series of questions that one might expect would be naturally
raised as a reaction to first reading about this theory, and immediate responses to each of these
questions.

\vspace{3ex}
\noindent
{\bf Question 1.} \emph{Why not always use the Bayesian approach to inference?}

\vspace{1ex}
As comments were already made in Section~\ref{sec3} regarding the flawed nature of two common
`objective' forms of Bayesian inference, let us consider the proposal of always making post-data
inferences about model parameters using the standard or subjective Bayesian paradigm.

It is clearly arguable that the main difficulty with the Bayesian paradigm is in choosing a prior
density function for the model parameters that adequately represents what was known about these
parameters before the data were observed. According to the definition of probability being adopted
in this paper, i.e.\ the definition outlined in detail in Bowater~(2018b) that was summarised in
Section~\ref{sec1}, carrying out this task in an unsatisfactory manner (which can reasonably be
regarded as often being unavoidable) is formally indicated by a low ranking being attached to the
external strength of the prior distribution function, under the assumption, which will be made from
now onwards, that the event $R(\lambda)$ is a given outcome of a well-understood physical
experiment (such as drawing a ball out of an urn of balls) and the resolution level $\lambda$ is
some value in the interval $[0.05,0.95]$.
In addition, it can be argued that, if we only apply Bayesian reasoning, then this assessment of
external strength should, in turn, generally result in a similar low ranking being attached to the
external strength of the posterior distribution function of the parameters that is based on the
prior distribution function concerned.

We can observe that it is often claimed that the choice of a prior distribution function is not
such an important issue if, over a set of `reasonable choices' for this distribution function, the
posterior distribution function to which it corresponds is not `greatly affected' by this choice.
However, it is difficult for such an argument to escape the issue that has just been raised, which,
in the present context, is the question of how externally strong should we regard any particular
posterior distribution function that corresponds to a prior distribution function that belongs to
the aforementioned set assuming that we can apply only Bayesian reasoning?
Furthermore, in response to the claim being considered, it can be argued that if, for example, we
had no or very little pre-data knowledge about the parameters of a given model, then the set of
`reasonable choices' for the prior density function of these parameters would need to be so diverse
that the corresponding posterior density function would indeed be very greatly affected by which
density function is chosen from this set.

Of course, if a prior density function can be found for a given set of parameters that is genuinely
considered to be a good representation of our pre-data knowledge about these parameters, then we
would naturally feel much less uneasy about the appropriateness of using the Bayesian method to
make inferences about the parameters concerned.
This is the reason why this method of inference is a critical component of the integrated framework
for data analysis that has been described in the present paper.

A more detailed discussion of the lines of reasoning that have just been presented can be found in
Bowater~(2017, 2018a, 2018b).
Moreover, it was also argued in detail in Bowater~(2018b) and Bowater~(2019a) that very high
rankings may be justifiably attached to the external strengths of fiducial distribution functions
derived by using the strong or moderate fiducial argument as part of the theory of organic fiducial
inference that was outlined in Section~\ref{sec4}, assuming that there was no or very little
pre-data knowledge about the parameters concerned over their permitted range of values.
Partially on the basis of this kind of reasoning, it could be argued furthermore that often, in
practice, similar high rankings should be attached to the external strengths of post-data
distribution functions derived using the type of bispatial inference described in
Section~\ref{sec5}, assuming that the scenario of interest specified in this earlier section is
strictly applicable.

\vspace{3ex}
\noindent
{\bf Question 2.} \emph{What about Lindley's criticism with regard to the incoherence of fiducial
inference?}

\vspace{1ex}
With reference to Fisher's fiducial argument, it was shown in Lindley~(1958) that, if the fiducial
density of a parameter $\theta$ that is formed on the basis of a data set $x$ is treated as a prior
density of $\theta$ in forming, in the usual Bayesian way, a posterior density of $\theta$ on the
basis of a second data set $y$, then, in general, this posterior density will not be the same as
the one that would be formed by repeating the same operation but with $y$ as the first data set,
and $x$ as the second data set, i.e.\ fiducial inference generally fails to satisfy a seemingly
reasonable coherency condition.

As a reaction to this, it can be remarked that fiducial inference, whether it is Fisher's version
of this type of inference, or the version outlined in the present paper, relies on pre-data
knowledge, or an expression of the lack of such knowledge, being incorporated into the inferential
process within the context of the observed data. Therefore, while it may be loosely acceptable, in
general, to apply a blanket rule such as the strong fiducial argument without concern for the data
actually observed, it is perhaps unsurprising that doing this could sometimes lead to the type of
phenomenon that has just been highlighted.
Also, the act of expressing pre-data knowledge is rarely going to be a completely 100\% precise act
no matter what paradigm of inference is adopted, therefore the door is always open for
inconsistencies in the inferential process such as the one identified in Lindley~(1958) that is
under discussion.
Furthermore, if indeed we are in a scenario where the coherency condition being considered is not
satisfied, then at least with respect to the type of fiducial inference outlined in the present
paper, i.e.\ organic fiducial inference, it would be expected that good approximate adherence to
this condition would usually be achieved providing that the data sets $x$ and $y$ referred to above
are at least moderately sized. In other words, it can be argued that the practical consequences of
the anomaly in question should generally be regarded as being quite small.

Observe that the same kind of anomaly is clearly also going to apply when post-data densities of
the parameters of a given model are constructed by relying in some way on the type of bispatial
inference that was described in Section~\ref{sec5}.
Similar arguments can be made, though, in response to the criticism being discussed with regard to
this type of situation as have just been presented.

Finally, we ought to mention an important issue that is related to this criticism.
In particular, if it is considered as being appropriate in a particular context to form a post-data
density function for the parameters of a given model by incorporating organic fiducial inference,
and possibly also bispatial inference, into the framework that has been detailed in the present
paper, then we may ask, would it not be best to use one or both of these methods of inference to
construct such a density function on the basis of a minimal part of the data set that has actually
been observed, and as a next step, use this density function as a prior density in analysing the
rest of the data under only the Bayesian paradigm?
Although, at first sight, this strategy may appear to be a reasonable one, it has the drawback that
post-data density functions constructed using organic fiducial inference on its own, or combined
with bispatial inference, may well be regarded as being less adequate representations of the
post-data uncertainty that is felt about the parameters concerned if they are based on a small
rather than a large amount of data.

For example, even if there was very little pre-data knowledge about a given parameter of interest
and the fiducial statistic $Q(x)$ is a sufficient statistic, it may be less appro\-priate to apply
the strong fiducial argument to make inferences about this parameter if the data set is small
rather than large.
Also, with regard to bispatial inference, there is of course generally less chance that the
one-sided P value in the hypothesis $H_{S}$ de\-fined by equation~(\ref{equ6}) or~(\ref{equ7}),
i.e.\ the value $F(t\,|\,\theta_j = \theta_{j0})$ or the value
$\Fprime(t\,|\,\theta_j = \theta_{j1})$, will be small if it is calculated on the basis of a small
rather than a large data set, and as a result, more chance perhaps that the interpretation of this
P value will be a little \linebreak complicated.

We are therefore led again to an issue that was discussed in the answer to Question 1 of this
section, in particular the question of whether we can justifiably attach a very high ranking to the
external strength of the prior density that forms the basis for carrying out the second step of the
type of strategy being considered and, if we can only apply Bayesian reasoning in this second
stage, whether we can justifiably attach a very high ranking to the external strength of the
posterior density that results from the whole analysis?

\vspace{3ex}
\noindent
{\bf Question 3.} \emph{If the choice of the fiducial statistic is not obvious, how should this
statistic be chosen?}

\vspace{1ex}
The definition of a fiducial statistic $Q(x)$ was given in Section~\ref{sec4}. As alluded to in
\linebreak this earlier section, if there is not a sufficient statistic for the unknown parameter
of interest that is a natural choice for the fiducial statistic, then a fairly general choice for
this latter statistic, which has a good deal of intuitive appeal, is the maximum likelihood
estimator of the parameter. Nevertheless, it would appear that more sophisticated cri\-teria for
choosing the fiducial statistic could be easily developed so that, in general, the effect of any
arbitrariness in the choice of this statistic could be assured as being negligible.
Such a development though will be left for future work.

\vspace{3ex}
\noindent
{\bf Question 4.} \emph{Can the results obtained from applying integrated organic inference depend
on the parameterisation of the sampling model?}

\vspace{1ex}
There are two key reasons why the parameterisation of the sampling model may possibly affect the
inferences made about population quantities of interest when applying integrated organic inference.
First, related to a point made in the answer to Question 2 of this section, it may be possible to
achieve a more representative expression of pre-data knowledge about the parameters of a model
using one parameterisation of the model rather than another.
In this case, it is fairly obvious that ideally, out of all possible parameterisations of the
model, the one should be chosen with regard to which the most representative expression of pre-data
knowledge about the parameters can be achieved.

The second reason why inferences may be possibly affected by model parameterisation is related to
the answer given to Question 3 of this section. In particular, it is that parameterisations may
exist with regard to which fiducial statistics $Q(x)$ or test statis\-tics $T(x)$ can be found that
make more efficient use of the information contained in the \linebreak data than those that can be
found with regard to other parameterisations.
However, it would be expected that, in general, this issue would not have more than a negligible
effect on post-data inferences made about quantities of interest, and where the effect of this
issue is more than negligible then, in the context of what was just discussed about the choice of
model parameterisation, there clearly should be a preference for those parameterisations that allow
fiducial statistics and test statistics to be chosen that make the best use of the information that
is in the data.

\vspace{3ex}
\noindent
{\bf Question 5.} \emph{In cases where the set of full conditional post-data densities referred to
in equation~(\ref{equ2}) are incompatible, how often, in practice, could we expect them to be
`approximately compatible'?}

\vspace{1ex}
Let us begin by clarifying that in interpreting this question it will be assumed that the full
conditional densities referred to in equation~(\ref{equ2}) would be described as being
`approximately compatible' if they were incompatible, but nevertheless it was possible to find a
joint density function of the parameters concerned such that these full conditional densities were
closely approximated by the full conditional densities of the given joint density.

In replying to the question just raised, let us first remember that examples were discussed in
Sections~\ref{sec12} to~\ref{sec7} of the present paper in which the Gibbs sampling method of
Section~\ref{sec2} was applied to determine a joint post-data density of the parameters of each of
the specific models of interest in these examples.
Also, various other examples of this kind were outlined in Bowater~(2018a, 2019a, 2019b).
In all of these examples, a justification was given as to why it would be reasonable to conclude
that if indeed the full conditional densities referred to in equation~(\ref{equ2}) are
incompatible, then they nevertheless should be approximately compatible.

However, let us take the opportunity to highlight two examples where the approximate compatibility
of the full conditional densities in equation~(\ref{equ2}) appeared to be less good than what was
seen to be generally the case in the examples of the type in question.
First, in an example in Bowater~(2018a) where organic fiducial inference was applied to the problem
of making post-data inferences about all the parameters of a bivariate normal distribution, a basic
simulation study showed that the full conditional densities referred to in equation~(\ref{equ2})
were clearly incompatible.
It could be argued, though, that the main reason for this was likely to be the fairly
unsophisticated normality assumptions that were made as part of this application of the method of
inference in question in order to approximate the full conditional fiducial densities for three of
the five parameters concerned, these three parameters being, in particular, the two population
variances and the correlation coefficient.
Second, although in an example in Bowater~(2019a) where organic fiducial inference was used to make
post-data inferences about all the parameters of a multinomial distribution, a justification was
given as to why the full conditional densities in equation~(\ref{equ2}) should be at least
approximately compatible, an additional (unreported) simulation study showed that in this example,
the full conditional densities in question often may not have this desirable property if the number
of trials (or in other words the number of observations) is very low and one or more of the
categories over which the multinomial distribution is defined contain no observations.
Nevertheless, the problem of making inferences about the parameters of a multinomial distribution
on the basis of limited data of this type when, as in the example being referred to, there is
assumed to be no or very little pre-data knowledge about the parameters concerned is generally a
difficult problem to solve using any paradigm of inference, see for example Berger, Bernardo and
Sun~(2015), and it is one that may well never have a completely satisfactory solution.

Finally, with regard to making inferences about the parameters $\theta$ of any given sam\-pling
model, it is important to bear in mind that, even if the full conditional densities re\-ferred to
in equation~(\ref{equ2}) fail to be at least approximately compatible, then nevertheless,
\linebreak as alluded to in Section~\ref{sec2}, they may well be considered as representing the
best information that is available for constructing the most suitable post-data density function
for the parameters concerned using the Gibbs sampling method outlined in this earlier section.

\vspace{2ex}
This concludes the discussion of the theory put forward in the present paper, i.e.\ integrated
organic inference (IOI). It is hoped that it will be appreciated that this theory modifies,
generalises and extends Fisherian inference, and naturally combines it with Bayesian inference in a
way that constitutes a major advance on the level of sophistication of either of these two older
schools of inference.

\vspace{5ex}
\noindent
{\bf References}

\begin{description}

\setlength{\itemsep}{1ex}

\item[] Bayes, T. (1763).\ An essay towards solving a problem in the doctrine of chances.\
\emph{Philosophical Transactions of the Royal Society}, {\bf 53}, 370--418.

\item[] Berger, J. O., Bernardo, J. M. and Sun, D. (2015).\ Overall objective priors.\
\emph{Bayesian Analysis}, {\bf 10}, 189--221.

\item[] Bowater, R. J. (2017).\ A defence of subjective fiducial inference.\ \emph{AStA Advances
in Statistical Analysis}, {\bf 101}, 177--197.

\item[] Bowater, R. J. (2018a).\ Multivariate subjective fiducial inference.\ \emph{arXiv.org
(Cornell University), Statistics}, arXiv:1804.09804.

\item[] Bowater, R. J. (2018b).\ On a generalised form of subjective probability.\ \emph{arXiv.org
(Cornell University), Statistics}, arXiv:1810.10972.

\item[] Bowater, R. J. (2019a).\ Organic fiducial inference.\ \emph{arXiv.org (Cornell University),
Sta\-tis\-tics}, arXiv:1901.08589.

\item[] Bowater, R. J. (2019b).\ Sharp hypotheses and bispatial inference.\ \emph{arXiv.org
(Cornell University), Statistics}, arXiv:1911.09049.

\item[] Brooks, S. P. and Roberts, G. O. (1998).\ Convergence assessment techniques for Markov
chain Monte Carlo.\ \emph{Statistics and Computing}, {\bf 8}, 319--335.

\item[] Chen, S-H. and Ip, E. H. (2015).\ Behaviour of the Gibbs sampler when conditional
distributions are potentially incompatible.\ \emph{Journal of \pagebreak Statistical Computation
and Simulation}, {\bf 85}, 3266--3275.

\item[] Cowles, M. K. and Carlin, B. P. (1996).\ Markov chain Monte Carlo convergence diagnostics:\
a comparative review.\ \emph{Journal of the American Statistical Association}, {\bf 91}, 883--904.

\item[] Efron, B. (1993).\ Bayes and likelihood calculations from confidence intervals.\
\emph{Bio\-met\-rika}, {\bf 80}, 3--26.

\item[] Gelfand, A. E. and Smith, A. F. M. (1990).\ Sampling-based approaches to calculating
marginal densities.\ \emph{Journal of the American Statistical Association}, {\bf 85}, 398--409.

\item[] Gelman, A. and Rubin, D. B. (1992).\ Inference from iterative simulation using multiple
sequences.\ \emph{Statistical Science}, {\bf 7}, 457--472.

\item[] Geman, S. and Geman, D. (1984).\ Stochastic relaxation, Gibbs distributions and the
Bayesian restoration of images.\ \emph{IEEE Transactions on Pattern Analysis and Machine
Intelligence}, {\bf 6}, 721--741.

\item[] Jeffreys, H. (1961).\ \emph{Theory of Probability}, 3rd edition, Oxford University Press,
Oxford.

\item[] Kass, R. E. and Wasserman, L. (1996).\ The selection of prior distributions by formal
rules.\ \emph{Journal of the American Statistical Association}, {\bf 91}, 1343--1370.

\item[] Lindley, D. V. (1958).\ Fiducial distributions and Bayes' theorem.\ \emph{Journal of the
Royal Statistical Society, Series B}, {\bf 20}, 102--107.

\item[] Metropolis, N., Rosenbluth, A. W., Rosenbluth, M. N., Teller, A. H. and Teller, E. (1953).\
Equation of state calculations by fast computing machines.\ \emph{Journal of Chemical Physics},
{\bf 21}, 1087--1092.

\end{description}

\end{document}